\documentclass[twocolumn,aps,prx,superscriptaddress,amsmath,amssymb]{revtex4-2}
\usepackage[T1]{fontenc}

\usepackage{siunitx}
\usepackage{gensymb,amssymb}
\usepackage[bottom]{footmisc}
\usepackage{lipsum}
\usepackage{graphicx}
\usepackage{dcolumn}
\usepackage{bm}
\usepackage{flushend}
\usepackage{xcolor}
\usepackage{lipsum}
\usepackage{multirow}
\usepackage{tcolorbox}
\usepackage{esvect}
\usepackage{textcomp}
\usepackage[shortlabels]{enumitem}
\usepackage{graphicx}
\usepackage{dcolumn}
\usepackage{bm}
\usepackage{amsmath}
\usepackage{algorithm}
\usepackage{algpseudocode}

\usepackage{soul}
\usepackage[normalem]{ulem}
\usepackage{array}

\usepackage[colorinlistoftodos]{todonotes}

\usepackage{hyperref}
\hypersetup{backref=true, 
  pagebackref=true
  urlcolor= blue, 
  linkcolor= blue, 
}

\newcommand{\mycomment}[1]{}

\begin{document}

\title{Physics-informed digital twin and onboard control of a brainbot for intelligent active matter}

\author{I.~Mammadli}
\affiliation{PULS, Institute for Theoretical Physics, FAU Erlangen-Nürnberg, 91058, Erlangen, Germany}
\author{P.~Shrestha}
\affiliation{Pattern Recognition Lab, Department of Computer Science, FAU Erlangen-Nürnberg, Erlangen, Germany}
\author{J.~Pande}
\email{Corresponding author: jayant.pande@flame.edu.in}
\affiliation{Department of Physical and Natural Sciences, FLAME University, Pune, India}
\author{F.~Novkoski}
\email{Corresponding author: filip.novkoski@fau.de}
\affiliation{PULS, Institute for Theoretical Physics, FAU Erlangen-Nürnberg, 91058, Erlangen, Germany}
\author{S.~Mohapatra}
\affiliation{PULS, Institute for Theoretical Physics, FAU Erlangen-Nürnberg, 91058, Erlangen, Germany}
\author{M.~Noirhomme}
\affiliation{GRASP, Institute of Physics B5a, University of Li\`ege, B4000 Li\`ege, Belgium}
\author{A.~Maier}
\affiliation{Pattern Recognition Lab, Department of Computer Science, FAU Erlangen-Nürnberg, Erlangen, Germany}
\author{N.~Vandewalle}
\affiliation{GRASP, Institute of Physics B5a, University of Li\`ege, B4000 Li\`ege, Belgium}
\author{A.-S.~Smith}
\affiliation{PULS, Institute for Theoretical Physics, FAU Erlangen-Nürnberg, 91058, Erlangen, Germany}
\affiliation{Group for Computational Life Sciences, Division of Physical Chemistry, Ru\dj{}er Bo\v{s}kovi\'c Institute, Zagreb 10000, Croatia}

\begin{abstract}
Establishing adaptive particles that sense their state, anticipate their evolution, and compute control inputs onboard has been a major challenge in non-equilibrium physics. We address this challenge by realizing an autonomous brainbot, building on a recently developed programmable bristlebot. First, we construct a physics-informed digital twin of the device, based on a kinematic model that reproduces measured trajectory statistics and generates long, statistically faithful synthetic trajectories. The kinematics forms the foundation for implementing onboard model predictive control (MPC), enabling autonomous trajectory tracking, demonstrated by accurate execution of a non-trivial target path. This provides a proof of principle for a brainbot that senses its state, predicts its evolution, and computes control inputs onboard, unlike conventional active particles with fixed motility, thereby transforming the brainbot into an agentic physical entity. By integrating physical modeling, data-driven parameter identification, and control into a unified framework, our approach provides a scalable platform for machine-learning-enabled multi-agent studies and lays the groundwork for intelligent, adaptive active matter.
\end{abstract}


\maketitle

\section{Introduction}\label{sec:introduction}

Active matter, unlike classical passive systems, consists of energy-consuming units operating far from equilibrium, enabling dynamic behaviors such as self-organization, collective motion, and synchronization. Several paradigmatic artificial designs have been proposed over the last decade to facilitate reproducible experimental study. These include systems such as active colloids that are capable of self-propulsion through mechanisms such as chemical reactions, light stimuli, and magnetic fields~\cite{Bechinger2016}, Quincke rollers that exhibit complex swirling motion~\cite{Liu2021,Garza2023}, and vibrated granular rods~\cite{Narayan2007}. Additionally, there are a variety of active asymmetric particles that move directionally due to imbalanced friction forces resulting from their shape or surface properties. One example of such systems is vibrobots, which are small particles placed on a vertically vibrating bed resulting in their horizontal motion~\cite{Scholz2018a,Scholz2018b}. While they function without internal processing, vibrobots can exhibit interesting self-organization and swarming behaviors~\cite{Caprini2024,Chen2024}. Recently, chiral vibrobots have been used to demonstrate the role of asymmetric interactions in active self-assembly~\cite{Caprini2024b}.

Another commonly used design of active particles, relying on an internal source of energy for motion generation, is bristlebots, such as the commercially available Hexbugs\texttrademark~\cite{Dauchot2019, Baconnier2022, Chor2023}. Bristlebots are equipped with flexible legs and an internal vibration motor powered by a battery, transforming vertical oscillation of the body into horizontal motion~\cite{Giomi2013}. This setup allows a considerably larger operational space, and has been used to study active particles in harmonic potentials~\cite{Dauchot2019}, to construct active solids~\cite{Baconnier2022} and elastoactive structures~\cite{Zheng2023}, and to mimic ant-like pheromone trail behavior by way of Hexbugs\texttrademark \ moving through fields of passive particles~\cite{Altshuler2024}.  

While bristlebots have been widely used, their locomotion became programmable and controllable only recently through the integration of ARM$^\text{\textregistered}$ microcontrollers into custom-built platforms, giving rise to so-called brainbots. These versatile, centimeter-sized devices are equipped with light, color, and magnetic field sensors, as well as infrared emitter–receivers~\cite{Noirhomme2025, Novkoski2025}. Powered by a lithium-ion battery providing $40$–$60$ minutes of operation, they exhibit motion patterns that can be tuned from ballistic to diffusive dynamics. Their trajectories depend sensitively on leg inclination and motor power~\cite{Noirhomme2025}. This behavior can be captured by a single parameter, $\eta$, which quantifies the similarity of a trajectory to pure spinning and varies systematically with driving strength and geometry, enabling continuous transitions between straight, spiral, and circular motion~\cite{Noirhomme2025}. However, the limited battery power restricts the exploration of long-term dynamics under complex driving protocols and the collective behavior of multiple units. This limitation could be overcome by developing a digital twin to systematically investigate such regimes computationally and experimentally validate only the most relevant scenarios. At present, such a digital twin is not available. 

Understanding the motion of bristlebots and brainbots is necessary to improve the control of their trajectories and enhance their performance and applicability. Physical models of bristlebot locomotion, based on the flexibility of the legs, have been introduced in one dimension~\cite{Giomi2013,Becker2014} and two dimensions~\cite{Majewski2017}. However, these have not been successfully combined with experimental data, nor do they account for the diversity of trajectories that such robots can exhibit~\cite{Noirhomme2025}. As such, a theoretical framework to capture the relevant kinematics has been missing. Due to their widespread use as active matter systems, endowing individual bristlebots with control capabilities would pave the way for them to become compact experimental platforms for intelligent active particles~\cite{Lowen2026}. Recent advances in the control algorithms tailored to resource-constrained microprocessors of the kind used on these bots have aided the development of such systems, enabling advanced control in real time~\cite{11246255, Nguyen2024, wikner2026realtimefeasibilityhighratempc}.

\begin{figure}[t!]
\includegraphics[width=0.99\linewidth]{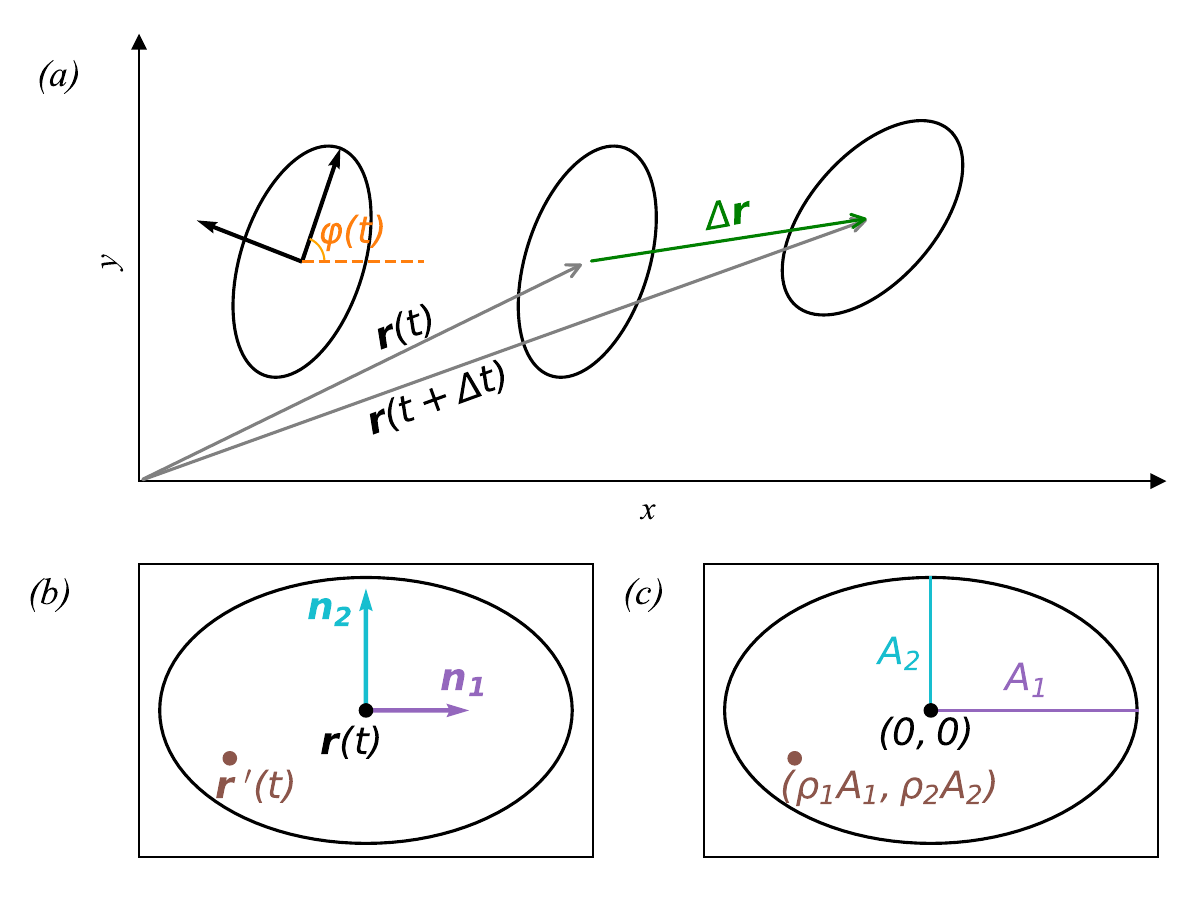}
\caption{(a) Position vector $\bm r(t)$ and orientation angle $\varphi(t)$ of a brainbot at time $t$, in the global reference frame. (b) Position vectors of the geometric center $\bm r$ of the brainbot and an arbitrary point $\bm r'$ in the laboratory frame, and the instantaneous unit vectors $\bm{\hat{n}_1}$ and $\bm{\hat{n}_2}$ along the major and minor axis, respectively. (c) Semi-major and semi-minor axis lengths $A_1$ and $A_2$, and the coordinates of the geometric center $(0,0)$ and an arbitrary point ($\rho_1 A_1, \rho_2 A_2$) in the body-fixed reference frame.}
\label{fig:bodyfixed2}
\end{figure}

In this work, we introduce a physics-based model that describes the various trajectories executed by a brainbot by decoupling the experimentally-observed translational, spinning, and orbital components of the motion. We also define an optimization procedure that fits the model to the experiments to generate relevant parameter sets that can describe the complex experimental trajectories with high accuracy, effectively building a physics-informed digital twin. We systematically evaluate the fidelity of the reconstruction as a function of the number of Fourier modes employed, enabling the generation of artificial trajectories whose $\eta$ descriptors quantitatively match the experimental templates on identical time scales. Furthermore, we use our digital brainbot to generate artificial trajectories on time scales far beyond experimentally-accessible ones. Notably, the distribution of $\eta$ that emerges from training on experimental data is preserved in this extrapolation. Moreover, ensembles of independently-generated long artificial trajectories maintain the same characteristic $\eta$ distributions, demonstrating the robustness and predictive consistency of our digital representation.

Because the digital brainbot reproduces behavior accurately across the full range of $\eta$, spanning diverse driving intensities and geometries, it provides a powerful tool for optimizing task-specific driving protocols. More broadly, it enables systematic exploration of complex physical scenarios through a synergistic interplay between experiment and \textit{in silico} realizations, made possible by the stability and fidelity of the digital twin established here.

Finally, our model of the brainbot, which relies on kinematics and doesn't require a complete resolution of the physics of the friction-based bristle-induced motion, is then employed to implement onboard predictive control of a single brainbot following a predefined trajectory. By relying on a lightweight solver designed for small-scale embedded systems, we enable a single brainbot to autonomously compute its control inputs to track a prescribed reference trajectory and make real-time adjustments in case of deviations. To the best of our knowledge, this is the first demonstration of on-board control for a bristlebot, underscoring the novelty of the approach and its relevance to the experimental realization of intelligent active matter.

\section{Kinematics of brainbot motion}\label{sec:kinematics}
\subsection{Analytical description of trajectories}\label{subsec:analytical}

Inspired by its experimental counterpart, our digital brainbot is assumed to move predominantly in two dimensions, here taken to be the $x$-$y$ plane. The brainbot is represented by an elliptical shape, whose geometric center is described by the two-dimensional position vector $\bm r(t)$ in the laboratory frame at time $t$. The orientation of the brainbot is described by the angle $\varphi(t)$ that the major axis of the brainbot makes with the horizontal $x$-axis at time $t$ (Fig.~\ref{fig:bodyfixed2}(a)).

Let $A_1$ and $A_2$ be the semi-major and semi-minor axis lengths of the brainbot ellipse, and let $\bm{\hat{n}_1}$ and $\bm{\hat{n}_2}$ be the unit vectors, in the laboratory frame, along the major and minor axes of the ellipse, given by
\begin{align}
    \bm{\hat{n}_1}(\varphi) &= (\cos\varphi, \ \sin\varphi),\label{eq:n1} \\
    \bm{\hat{n}_2}(\varphi) &= (-\sin\varphi, \ \cos\varphi).
    \label{eq:n2}
\end{align} 
Let $\bm r'$ denote an arbitrary point in the plane (inside or outside the ellipse) in the laboratory frame, and let its coordinates be $\rho_1 A_1$ and $\rho_2 A_2$ in the reference frame of the brainbot, where $\rho_1$ and $\rho_2$ are scale factors. If $\bm r'$ is on or inside the ellipse, then the $\rho$'s must lie between $-1$ and $1$, and must in fact satisfy the condition
\begin{equation}    \rho_1^2+\rho_2^2\leq 1.
\end{equation}

The coordinates of the geometric center $\bm r$ and the arbitrary point $\bm r'$ are related by
\begin{equation}
    \bm r' = \bm{r} + \rho_1 A_1 \bm{\hat{n}_1} + \rho_2 A_2 \bm{\hat{n}_2},
    \label{eq:r_rc_rel}
\end{equation}
while their velocities ($\bm v'$ at $\bm r'$ and $\bm v$ at $\bm r$) are related by
\begin{equation}
    \bm v' = \bm v + \bm \omega \times (\bm r' - \bm r),
    \label{eq:vfull}
\end{equation}
where $\bm \omega \equiv (0,0,\omega)$ is the angular velocity vector.

Let ${\bm r_{\rm c}}(t)$ denote the instantaneous center of rotation of the brainbot, i.e., the point at a given instant at which the velocity $\bm v'$ is zero. Then
\begin{equation} \label{eq:icr_calc}
\begin{aligned}
    {\bm r_{\rm c}}(t) \equiv \left(x - \frac{v_y}{\omega}, y + \frac{v_x}{\omega}\right),
\end{aligned}
\end{equation}
where $(x, y)$ and $(v_x, v_y)$ are the components of $\bm r$ and $\bm v$, respectively. At a given moment, the center of rotation of the brainbot may lie inside or outside the brainbot.

With these definitions, we introduce a curvature parameter $\eta$ to quantify the similarity of an observed trajectory to spinning motion
\begin{equation}
    \eta = \frac{|\omega| \  |\bm r - \bm r_{\rm c}|}{|\bm v|},
    \label{eq:eta}
\end{equation}
As discussed in our previous work~\cite{Noirhomme2025}, this parameter is useful to characterize the different trajectories,
with values of $\eta$ close to $0$ indicating linear motion and values close to $1$ indicating spinning motion.

\begin{figure*}
\includegraphics[width=1.0\textwidth]{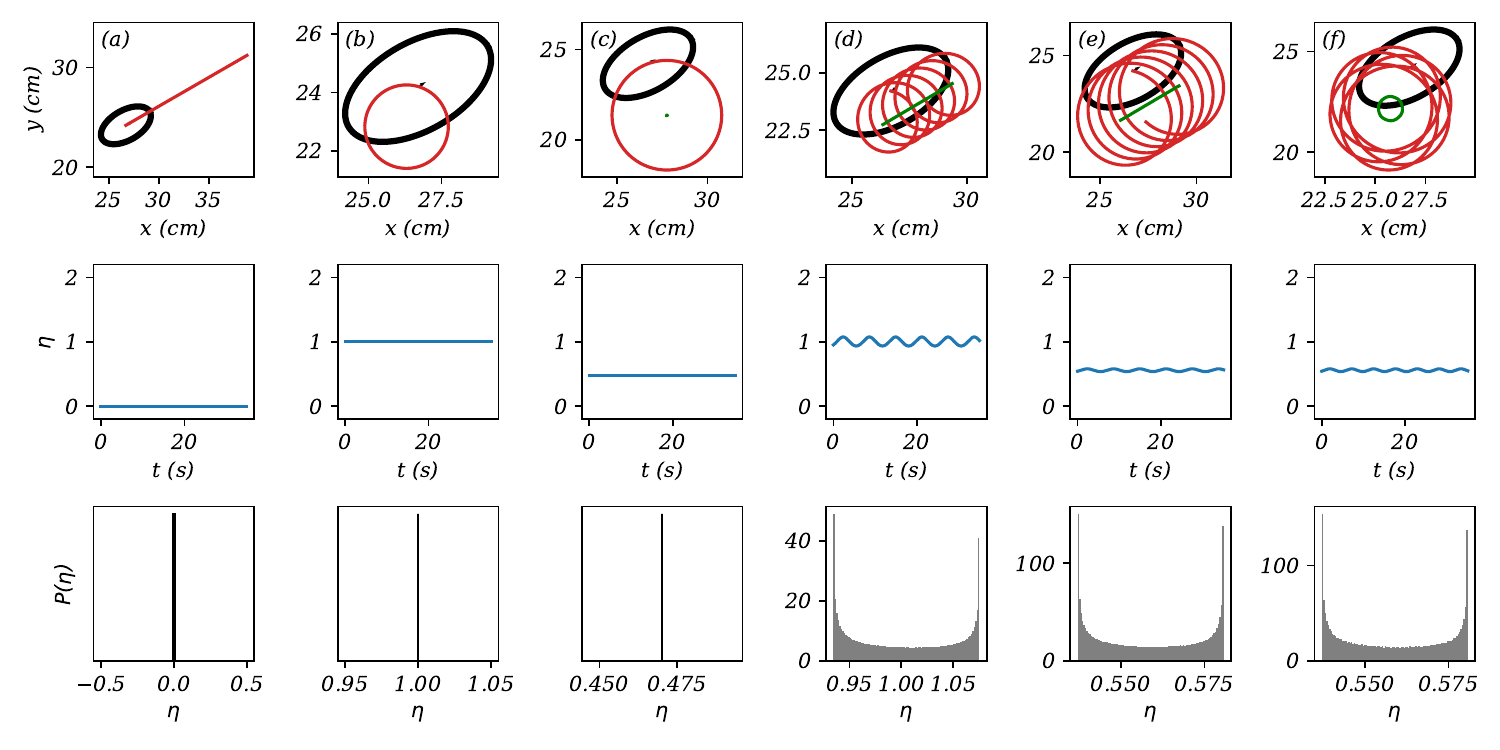}
\caption{Top row: Deterministic trajectories of the geometric center (shown in red) and the instantaneous center of rotation $\bm r_{\rm c}$ (shown in green) of a brainbot, found using the kinematic model. The black ellipse in each panel indicates the starting position of the brainbot. 
Middle row: The values of the $\eta$ parameter corresponding to the different trajectories shown in the top row. For purely linear, spinning, and orbital trajectories, the $\eta$ values are constant, while for other trajectories, $\eta$ is sinusoidal. Bottom row: The probability distributions of the different $\eta$ values attained in the corresponding plots in the middle row.   
 To get each distribution, the trajectory was calculated for $200$ seconds.  For visualization purposes, the plotted trajectories cover only the first $30$ seconds.}
\label{fig:sim1} 
\end{figure*}

To understand the relationship between the distribution of $\eta$ along trajectories and the overall properties of brainbot motion, we study some basic driving protocols. Specifically, we present a kinematic model that assigns to the brainbot a translational velocity $\bm v_\text{c}$ that is held constant either in the body-fixed frame or the laboratory frame.

\subsubsection{$ v_\text{c}$ constant in the body-fixed frame}\label{subsubsec:vcbody}
We first explore the case of constant $\bm v_\text{c}$ always applied at the point $(\rho_1 A_1, \rho_2 A_2)$ in the body-fixed frame, where $\rho_1$ and $\rho_2$ are constant. This may result, for example, from an internal driving force acting along the long axis of the ellipse. In the absence of an internal rotation, this trivially results in motion along a straight line (Fig.~\ref{fig:sim1} (a)). Introducing a constant $\omega$, which seems to be intrinsic to the experimental design, gives $\bm v_\text{c}$ of the form
\begin{equation}
\bm v_\text{c} = u_{1} \bm{\hat{n}_1} +u_{2} \bm{\hat{n}_2},
\end{equation}
where $u_{1}$ and $u_{2}$ are the constant velocities in the $\bm{\hat{n}_1}$ and $\bm{\hat{n}_2}$ directions, respectively. Using Eq.~(\ref{eq:vfull}) we obtain
\begin{align}\label{eq:drdt}
\dfrac{\mathrm d}{\mathrm dt}\bm r = & u_{1} \bm{\hat{n}_1} +u_{2} \bm{\hat{n}_2} - \bm \omega\times\left(\rho_1 A_1 \bm{\hat{n}_1} + \rho_2 A_2 \bm{\hat{n}_2}\right) \nonumber\\
 = & \left[\sin\varphi\left(\omega\rho_1 A_1-u_2 \right) + \cos\varphi\left(\omega\rho_2 A_2 + u_1\right)\right]\bm{\hat{i}} \nonumber\\&+ \left[\sin\varphi\left(\omega\rho_2 A_2 + u_1\right) + \cos\varphi\left(- \omega\rho_1 A_1 + u_2 \right)\right]\bm{\hat{j}}
\end{align}
where $\bm{\hat{i}}$ and $\bm{\hat{j}}$ denote the unit vectors along the $x$- and $y$-directions, respectively, in the laboratory frame. Integrating with respect to $t$ (with $\mathrm dt = \mathrm d\varphi/\omega$), we get
\begin{align}\label{eq:rfull}
\bm r(t) = \bm r(0) + &\left\{\left(\rho_2 + \dfrac{u_1}{A_2 \omega}\right)A_2\left[\sin\varphi(t)-\sin\varphi(0)\right]\right.\nonumber\\
&+\left.\left(-\rho_1 + \dfrac{u_2}{A_1 \omega}\right)A_1\left[\cos\varphi(t)-\cos\varphi(0)\right]\right\}\bm{\hat{i}}\nonumber\\
+ &\left\{\left(-\rho_1 + \dfrac{u_2}{A_1 \omega}\right)A_1\left[\sin\varphi(t)-\sin\varphi(0)\right]\right.\nonumber\\
&-\left.\left(\rho_2 + \dfrac{u_1}{A_2 \omega}\right)A_2\left[\cos\varphi(t)-\cos\varphi(0)\right]\right\}\bm{\hat{j}}.
\end{align}
Defining the modified scale factors $\rho_1' \equiv \rho_1 - u_1/(A_2 \omega)$ and $\rho_2' \equiv \rho_2 + u_1/(A_2 \omega)$, 
we obtain

\begin{align}\label{eq:rsimple}
\bm r(t) = \bm r_\text{const} - R\left\{\cos\left[\varphi(t)+\alpha\right] \bm{\hat{i}}+\sin\left[\varphi(t)+\alpha\right] \bm{\hat{j}}\right\},
\end{align}
where the constants $\bm r_\text{const}$, $R$ and $\alpha$ are defined by

\begin{align}
\bm r_\text{const} &= \bm r(0) -\left[\rho_2'A_2\sin\varphi(0) - \rho_1'A_1\cos\varphi(0)\right]\bm{\hat{i}}\nonumber\\
& \ \ \ +\left[\rho_1'A_1\sin\varphi(0) + \rho_2'A_2\cos\varphi(0)\right]\bm{\hat{j}},\label{eq:rconst}
\end{align}
\begin{align}
R &= \sqrt{\rho_1'^2 A_1^2  + \rho_2'^2 A_2^2}\label{eq:R}
\end{align}
and
\begin{align}\label{eq:alpha}
\alpha &= \cos^{-1}\left(\dfrac{\rho_1' A_1}{\sqrt{\rho_1'^2 A_1^2 + \rho_2'^2 A_2^2}}\right) \\&=
\sin^{-1}\left(\dfrac{\rho_2' A_2}{\sqrt{\rho_1'^2 A_1^2 + \rho_2'^2 A_2^2}}\right).
\end{align}

Eq.~(\ref{eq:rsimple}) shows that for a fixed $\omega$ and $\bm v_\text{c}$ constant in the body-fixed frame, the center of the brainbot executes a circular trajectory in the laboratory frame (Fig.~\ref{fig:sim1} (b)), with its center at $\bm r_\text{const}$ and its radius given by $R$. 

If $\bm v_\text{c} = \bm 0$, then the brainbot undergoes purely spinning motion (Fig.~\ref{fig:sim1} (b)), centered at the point $(\rho_1 A_1, \rho_2 A_2)$ in the body-fixed frame (which in this case is also fixed in the laboratory frame, at $\bm r_\text{const}$ given by Eq.~(\ref{eq:rconst})). In this case, the radius of the spinning trajectory is $\sqrt{\rho_1^2 A_1^2 + \rho_2^2 A_2^2}$. 

For a non-zero $\bm v_\text{c}$, the radius of the circular trajectory can be larger or smaller than the spinning radius (Fig.~\ref{fig:sim1} (c)). We call this orbital motion. In this case, depending on the signs of $u_1$ and $u_2$, the center of rotation may move inward (toward the geometric center of the brainbot) or outward, and can lie outside the brainbot.

\subsubsection{$ v_\text{c}$ constant in the laboratory frame}\label{subsubsec:vclab}
We now consider the case when $\bm v_\text{c}$ is a constant vector in the laboratory frame, and is again applied at the fixed point in the brainbot defined by $(\rho_1 A_1, \rho_2 A_2)$ in the body-fixed frame, where $\rho_1$ and $\rho_2$ are constant. This corresponds to subjecting the device to a constant external force. Under such circumstances, and with constant $\omega$, $\bm v_\text{c}$ is sinusoidal in the body-fixed frame (see Sec.~I of Supplementary Information~\cite{supplementary}).

The analogue of Eq.~(\ref{eq:rsimple}) in this case is
\begin{align}\label{eq:rsimplev2}
\bm r(t) = \bm r_\text{const} + \bm v_\text{c}t - R&\left\{\cos\left[\varphi(t)+\alpha\right] \bm{\hat{i}}\right. \nonumber\\
&+\left.\sin\left[\varphi(t)+\alpha\right] \bm{\hat{j}}\right\},
\end{align}
where $\bm r_\text{const}$, $R$ and $\alpha$ are as defined in Eqs.~(\ref{eq:rconst})-(\ref{eq:alpha}), except with $\rho_1'$ and $\rho_2'$ replaced by $\rho_1$ and $\rho_2$, respectively.

Eq.~(\ref{eq:rsimplev2}) shows that when $\bm v_\text{c}$ is kept fixed in the laboratory frame, the geometrical center of the brainbot undergoes a combination of translational motion (at constant velocity $\bm v_\text{c}$) and circular motion (which is a combination of spin and orbital motion, as in section~\ref{subsubsec:vcbody}). This results in helical trajectories (Fig.~\ref{fig:sim1} (d),(e)).

\subsection{Classification of trajectories}

The different forms of $\bm v_\text{c}$ discussed in sections~\ref{subsubsec:vcbody} and \ref{subsubsec:vclab} allow us to reproduce the various autonomous trajectories executed by a brainbot. The different fundamental modes of motion are discussed in Table~\ref{tab:cases} with the corresponding simulated trajectories shown in Fig.~\ref{fig:sim1}, top row.

\begin{table}[ht!]
\small 
\setlength{\tabcolsep}{3pt} 
\renewcommand{\arraystretch}{1.3} 
\begin{tabular}{|p{1.1cm}|p{1.75cm}|p{1.15cm}|p{3cm}|}
  \hline  \textbf{Fig.~\ref{fig:sim1} panel} & \textbf{Trajectory} & $\bm{\omega}$ \textbf{(rad/s)} & $\bm v_\text{c}$ \textbf{in the body-fixed frame (cm/s)} \\
  \hline
  (a) & Linear & 0   & 0.4 \\
  \hline
  (b) & Spinning & -1.0 & 0  \\
  \hline
  (c) & Orbital & -1.0 & (2, 0.5) \\
  \hline
  (d) & Helical & -1.0 & $0.1(\cos\omega t,-\sin\omega t)$ \\
  \hline
  (e) & Orbital + Helical & -1.0 & $(0.08,0.08)+0.1(\cos\omega t,-\sin\omega t)$ \\
  \hline
  (f) & Unobserved trajectory (frequency $\nu \neq \omega)$ & -1.0 & $(0.08,0.08)+0.1(\cos\nu t,-\sin\nu t)$ \\
  \hline
\end{tabular}
\caption{Input velocities for the simulation trajectories shown in Fig.~\ref{fig:sim1}. The cases $\omega = 0$ and $v_\text{c} = 0$ result in purely linear and purely spinning motion, respectively. In cases (d) and (e), the velocity $v_\text{c}$ is sinusoidal in the body-fixed frame (and constant in the laboratory frame, see Sec.~I of Supplementary Information for details~\cite{supplementary}) and is defined using the oscillation frequency $\omega$. In case (f), a different frequency $\nu = -1.2 \text{ rad/s} \neq \omega$ is used to define $v_\text{c}$, resulting in a closed-loop trajectory that has not been observed in the experiments.}
\label{tab:cases}
\end{table}

\begin{figure*}[t!]
\includegraphics[width=0.8\textwidth, trim=1.5cm 0.24cm 2cm 0.5cm, clip]{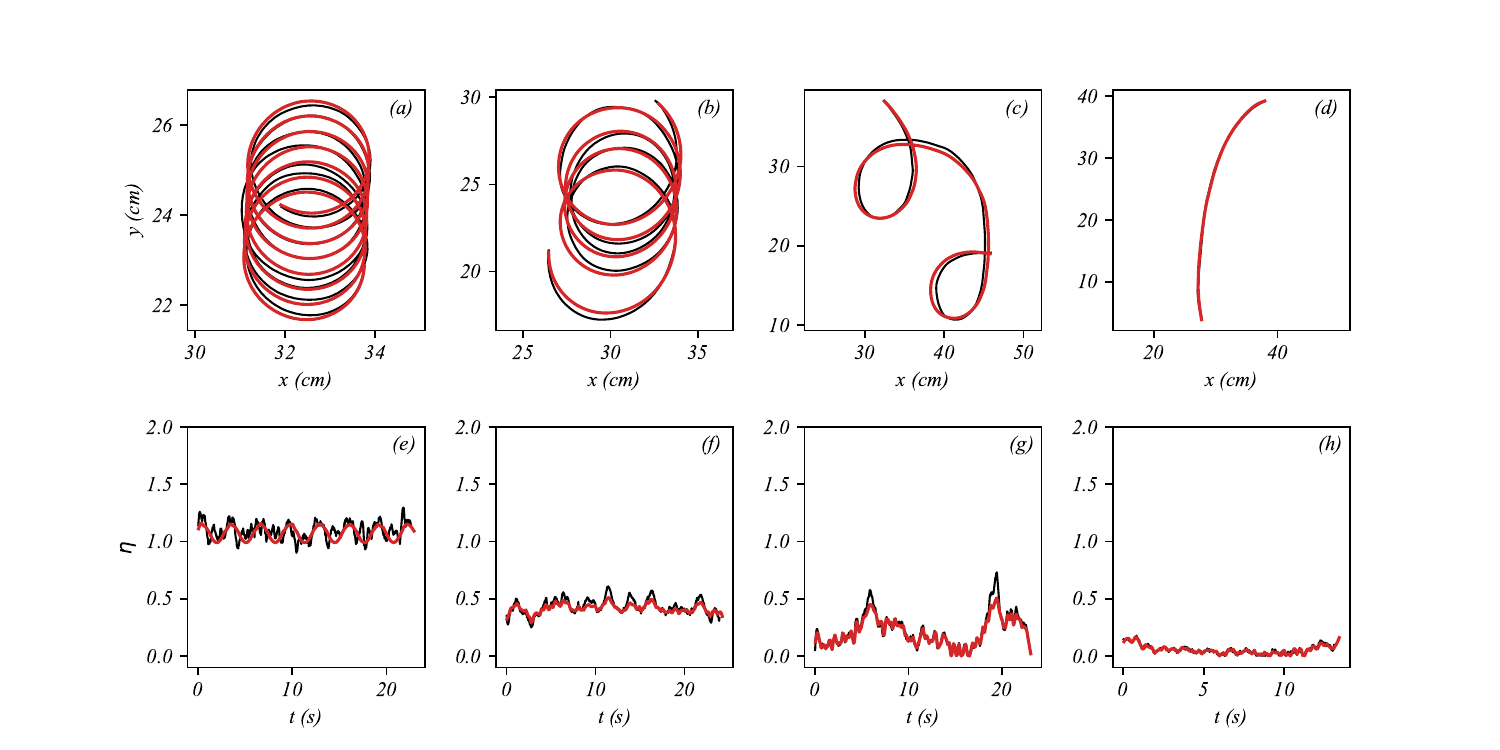}
\caption{Results of simulation (in red) compared to experimental data (in black). The upper row shows the trajectories, while the lower row shows the corresponding values of the $\eta$ parameter for each trajectory as a function of time. Values of $\eta$ close to $1$ indicate circular trajectories, while values close to $0$ indicate linear trajectories~\cite{Noirhomme2025}. The experimental parameters (leg angle $\alpha_\text{leg}$ and effective motor voltage $V_\text{E}$) in panels (a) to (d) are: (a) $\alpha_\text{leg} = 15 \degree, \ V_\text{E} = 2.7 \  \text{V}$, (b) $\alpha_\text{leg} = 15 \degree, \ V_\text{E} = 3.0 \ \text{V}$, (c) $\alpha_\text{leg} = 25 \degree, \ V_\text{E} = 2.1 \ \text{V}$, (d) $\alpha_\text{leg} = 5 \degree, \ V_\text{E} = 2.1 \ \text{V}$.}
\label{fig:traj_eta} 
\end{figure*}

To understand how the above-discussed modes of motion (top row in Fig.~\ref{fig:sim1}) correspond to the classification parameter $\eta$, we explore the time evolution of $\eta$ (second row in Fig.~\ref{fig:sim1}) and its associated distribution (third row in Fig.~\ref{fig:sim1}).
Linear and circular trajectories (the latter including both spinning and orbital trajectories) are characterized by a single value of $\eta$, resulting in a delta-function distribution $P(\eta)$. For linear and spinning motion $\eta$ equals $0$ and $1$, respectively, while for orbital motion $0 < \eta < 1$. 

For the helical (Fig.~\ref{fig:sim1} (d),(e)) and more complex trajectories (Fig.~\ref{fig:sim1} (f)), $\eta$ varies with time following one sinusoidal mode. Consistent with the probability distribution of any sinusoidal function with binning, in these cases $\eta$ has a probability distribution that is peaked at the extreme values of $\eta$ and is at its minimum at the mean value of $\eta$ (Fig.~\ref{fig:sim1} (d)-(f), third row).

The experimental trajectories are inherently more complex, but our analysis shows that they can be represented as a superposition of the principal modes of motion identified here. For example, one can extract $\omega(t)$ from the experimental trajectory and perform a Fourier decomposition to different orders (see Fig.~S1 and Sec.~III of Supplementary Information~\cite{supplementary}).
For the sample trajectory shown in Fig.~S1 of Supplementary Information~\cite{supplementary}, we find that at least $10$ modes are needed to get a reasonable distribution of $\eta$, while about $70$ modes are required for an accurate representation of the measured trajectory.

The complexity of the experimental trajectories suggests that for a digital representation of the brainbot to be faithful to the experiments, we must choose an efficient approach for removing high-frequency noise in the data that provides an optimal trade-off between the descriptive accuracy achieved and the computational overhead required. 

\section{Digital brainbot}\label{sec:simulations}

\subsection{Forward kinematic framework}\label{sec:simulations}
The establishment of a digital twin of the brainbot requires the development of a numerical framework that guarantees adherence of the bot’s motion to the underlying physical constraints. Consequently, the principles of the motion formalized above are cast into a forward kinematic model. This requires numerically integrating ${\rm d}{\bm r}/{\rm d}t$, like in Eq.~(\ref{eq:drdt}). A pseudo-code for the digital brainbot is given in Algorithm~\ref{alg:bbotSim}. Details of the numerical integration scheme used, including the rotation matrix $\bm M(\bar \omega \Delta t)$ and the mean velocity $\bar{\bm v}_\text{c}$, are provided in Sec.~II of Supplementary Information~\cite{supplementary}.

\subsection{Offline parametric identification model}\label{subsec:reproducing_exp}

To accurately represent the brainbot in the digital form, the forward kinematic framework must be complemented with time-dependent $\bm v_\text{c}$ and $\omega$. These parameters must be inferred from the experimental data. To this end, we rely on a gradient descent method to minimize the overall least-squares error between the experiments and the kinematic framework across the trajectories, for given geometric and driving conditions. We consider the process complete when the variance in the parameters stabilizes, which indicates that the digital twin effectively reflects and generalizes the brainbot’s actual physical behavior.

For this purpose, we use a trajectory dataset that comprises the position $(x,y)$ and the orientation $\varphi$ of the brainbot’s geometric center at each time step. Each trajectory lasts approximately $20$ seconds and contains about $800$ data points. To suppress high-frequency noise, we apply a Savitzky–Golay filter to the trajectories~\cite{savgol}, following the procedure described in Ref.~\cite{Noirhomme2025}.

From these processed trajectories, $\omega(t)$ can be obtained by numerically differentiating $\varphi(t)$ with respect to time. This sequence is used to generate a histogram of $\omega$ values for each trajectory across the entire run of the experiment and then fit the resulting distribution of $\omega$ to a skew-normal distribution. These distributions are stored for each driving and geometry condition of the brainbot.

To extract $\bm v_\text{c}$, we assume it to be a sum of the velocity profiles in sections~\ref{subsubsec:vcbody} and \ref{subsubsec:vclab}. That is, in the laboratory frame we assume  $\bm v_\text{c}$ to be a sum of constant and sinusoidal terms, namely,
\begin{align}
    \bm v_\text{c}(t) &= \left[u_{11} + u_{12}\cos(\varphi(t) + \alpha_1)\right] \bm{\hat{n}_1} \nonumber\\
    &+ \left[u_{21} + u_{22}\cos(\varphi(t) + \alpha_2)\right] \bm{\hat{n}_2}.
    \label{eq:icr}
\end{align}

The six parameters $\bm u \equiv (u_{11}, u_{12}, u_{21}, u_{22}, \alpha_1, \alpha_2)$ are found by minimising $ |\bm r(t) - \bm f_s (t; \bm u)|^2$ over $\bm u$ (least-squares), across all the $800$ data points in each trajectory, using the explicit $\omega(t)$ sequence. Here $\bm r(t) \equiv (x(t), y(t))$ denotes the trajectory points and $\bm f_s(t; \bm u)$ is the simulator that integrates the $\bm u$ parameters. 

Providing the six parameters to the kinematic framework completes the digital representation of the brainbot. 
We compare four experimental and synthetic trajectories (Fig.~\ref{fig:traj_eta} (a)-(d)) with examples progressively going from circular to linear to cover the full range of possible driving protocols and bot geometries. This is reflected in the $\eta$-parameter values corresponding to these trajectories, shown in panels (e)-(h) in Fig.~\ref{fig:traj_eta}, ranging from around $1$ to $0$ ~\cite{Noirhomme2025}. The very good agreement between the experimental and the simulated curves indicates that the form of $\bm v_\text{c}$ in Eq.~(\ref{eq:icr}) is justified and that the parametric identification model is robust.

\begin{algorithm}[H]
\caption{Digital Brainbot Kinematic Framework}
\begin{algorithmic}[1]
\State Initialise Brainbot $(x^{(0)}, y^{(0)}, \varphi^{(0)})$
\vspace{0.5em}
\State Prescribe $\bm v_\mathrm{c}(t)$ and $\omega(t)$
\vspace{0.5em}
\State $\bm{\hat{n}_1}^{(0)} \gets \bm{\hat{n}_1}(\varphi^{(0)})$ and $\bm{\hat{n}_2}^{(0)} \gets \bm{\hat{n}_2}(\varphi^{(0)})$
\vspace{0.5em}
\State $\rho_1\gets-0.374$, $\omega^{(0)}\gets \omega(0)$, $\bm v_\mathrm{c}^{(0)} \gets \bm v_\mathrm{c}(0)$
\vspace{0.5em}
\For{$k = 0 $ to $N_\mathrm{steps}-1$}
    \vspace{0.5em}
    \State $\omega^{(k+1)} \gets \omega((k+1)\Delta t)$,\;\; $\bm v_\mathrm{c}^{(k+1)} = \bm v_\mathrm{c}((k+1)\Delta t)$
    \vspace{0.5em}
    \State $\bar \omega = 0.5(\omega^{(k)} + \omega^{(k+1)})$
    \vspace{0.5em}
    \If{$\operatorname{sign}(\bar \omega^{(k+1)}) \neq \operatorname{sign}(\bar \omega^{(k)})$}
        \vspace{0.5em}
        \State   $\rho_2^{(k)}  \gets 0.661 \cdot \mathrm{sign}(\bar \omega )$
        \vspace{0.5em}
        \State  $\bm r_\mathrm{c}^{(k)} \gets \bm r^{(k)} + \rho_1^{(k)}A_1 \bm{\hat{n}_1}^{(k)} +\rho_2^{(k)}A_2 \bm{\hat{n}_2}^{(k)}$
        \vspace{0.5em}
    \EndIf
    \vspace{0.5em}
    \State $\varphi^{(k+1)} \gets \varphi^{(k)}+\bar \omega^{(k)}\Delta t$
    \vspace{0.5em}
    \State $\bm{\hat{n}_{1,2}}^{(k+1)} \gets \bm{\hat{n}_{1,2}}(\varphi^{(k+1)})$
    \vspace{0.5em}
    \State $\bar{\bm  v}_\mathrm{c} = 0.5(\bm v^{(k)}_\mathrm{c}+\bm v^{(k+1)}_\mathrm{c})$
    \vspace{0.5em}
    \State $\bm r_\mathrm{c}^{(k+1)} \gets \bm r_\mathrm{c}^{(k)}+\bar{\bm v}_\mathrm{c}\Delta t$
    \vspace{0.5em}
     \State $\bm r^{(k+1)} \gets \bm M(\bar \omega \Delta t)(\bm r^{(k)} - \bm r_\mathrm{c}^{(k)})+\bm r_\mathrm{c}^{(k+1)}$
     \vspace{0.5em}
\EndFor
\end{algorithmic}
\label{alg:bbotSim}
\end{algorithm}

\subsection{Applying the brainbot digital twin to generate long synthetic trajectories}\label{subsec:generating_traj}

\begin{figure*}
    \centering
    \includegraphics[width=0.8\textwidth, trim=1.6cm 2.8cm 2.5cm 3.cm, clip]{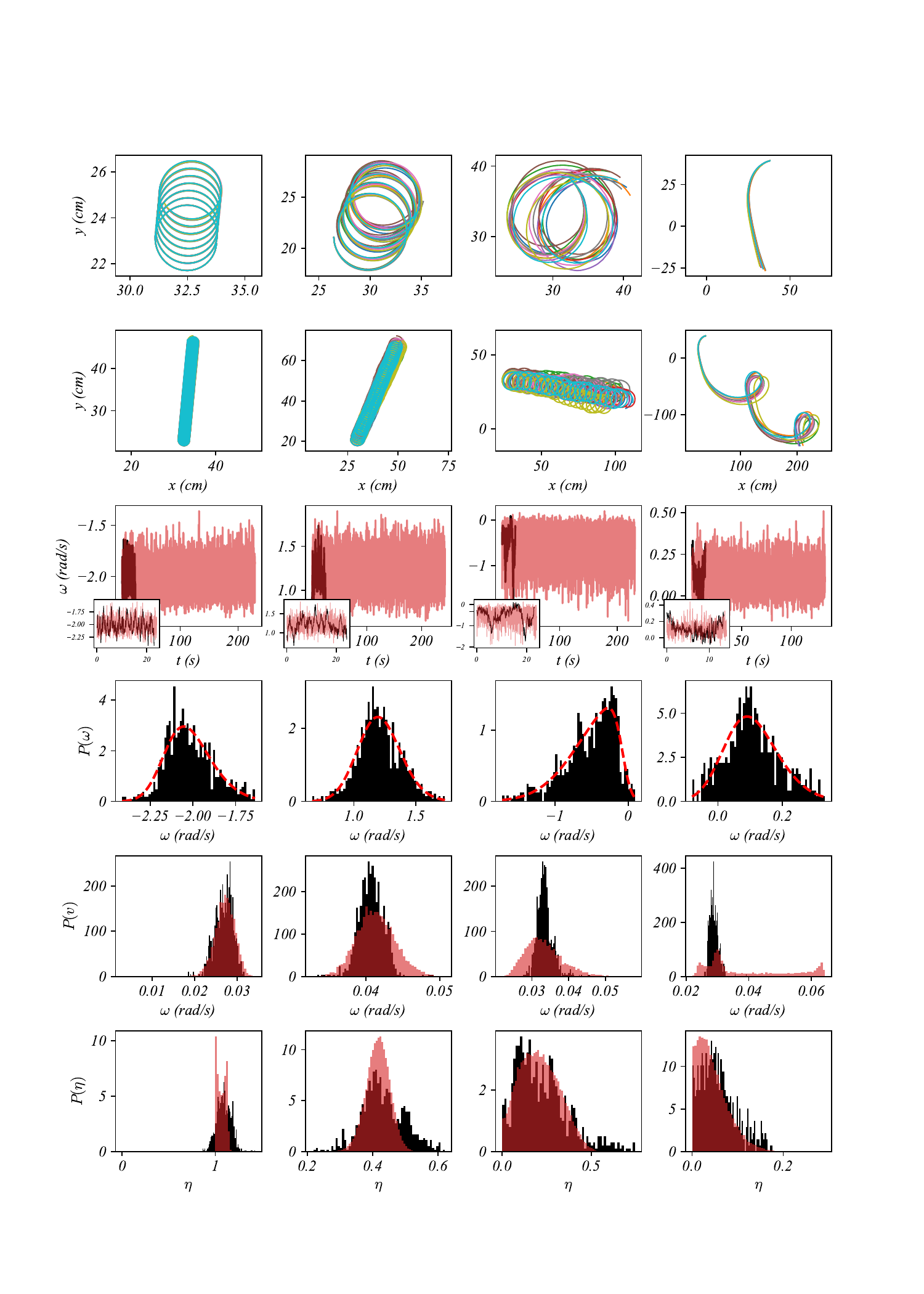}
    \caption{
    Top row: Synthetic trajectories similar to the four experimental trajectories shown in Fig.~\ref{fig:traj_eta}, and run for the same time duration. Second row: The same respective trajectories but now for simulation time steps that are ten times longer than the experimental duration. Third row: Variation in $\omega$ with time, from the experiments (black curves) and the long simulations (red curves), with the insets showing the comparison when the time durations are kept the same for the simulations and the experiments. Fourth row: Histograms (in gray) of $\omega$ values obtained from the simulations, and the resulting skew-normal distribution function (shown by the dashed red line), from which  $\omega$ values are picked to obtain the trajectories in the top row. Fifth row: Histograms of $v$ from the experiments (in gray) and the simulations (in red). Bottom row: Histograms of $\eta$ values from the experiments (in black) and from the simulations (in red).}
    \label{fig:omega_dist} 
\end{figure*}

In the previous section, synthetic trajectories arose from the parametric optimization procedure used to construct the digital twin. Here, we employ the fully trained digital twin to generate ensembles of synthetic trajectories, using the training data shown in Fig.~\ref{fig:traj_eta} as reference. For each parameter set, ten independent trajectories were generated. The complete ensemble is shown in Fig.~\ref{fig:omega_dist}, where the top row displays trajectories of the same duration as the experimental template, while the second row presents trajectories that are ten times longer.

For each parameter set and for each trajectory, we use the values of $\bm u$ as found in Section~\ref{subsec:reproducing_exp}, while the values of $\omega$ are drawn from the skew-normal distribution (dashed red line in Fig.~\ref{fig:omega_dist}, fourth row), yielding sequences for $\omega$ shown in the third row of Fig.~\ref{fig:omega_dist}, with the dark curves showing the training set and the red curves the generated set. The main panels in this row are for long times (ten times as long as the experiments) while the insets show the $\omega$ plots for short times (as long as the experiments).

\begin{figure*}[t!]
\includegraphics[width=0.99\linewidth]{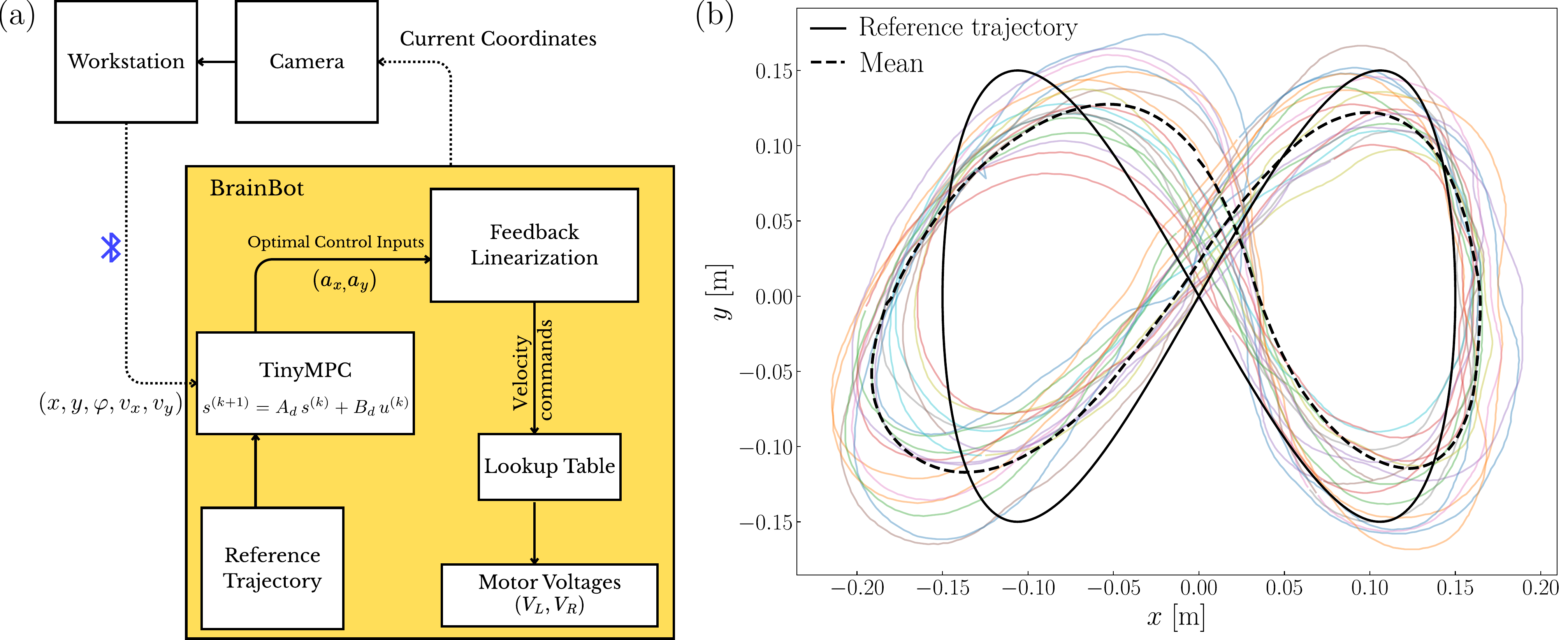}
\caption{Left: Schematic diagram of the model predictive control implementation for a brainbot, consisting of the onboard control based around TinyMPC (yellow) and the external positioning system, relying on a camera and tracking software, with the two communicating via Bluetooth. Right: Trajectory of a brainbot controlled through TinyMPC as it executes a lemniscate (figure eight), using a reference trajectory (solid black). Each of the differently colored lines represents individual cycles of the lemniscate, with the mean trajectory represented by the dashed black line.}
\label{fig:fig8}
\end{figure*}

Visual inspection of the synthetic trajectories demonstrates the robustness and accuracy of the brainbot's digital twin. Each reinitialization produces a new trajectory that is consistent with the distributions of the linear velocity and $\eta$ across the phase diagram (fifth and last rows in Fig.~\ref{fig:omega_dist}, respectively), while the distribution of $\omega$ is maintained by construction. Even a significant extrapolation in time keeps the statistical properties of the trajectories intact (as evidenced by the distributions of $v$ and $\eta$). At the same time, it expands the observed morphology of the trajectories: for instance, the apparently linear trajectory in Fig.~\ref{fig:omega_dist} (d) becomes helical when extended for a long time. This shows the digital twin to be a faithful and informative extension of the experimental brainbot, while being computationally efficient.  

\section{Implementation of the kinematic framework on a physical bot}


We now exploit the kinematic model to control a physical brainbot~\cite{Novkoski2025}. To this end, we employ model predictive control (MPC)~\cite{kouvaritakis2016model}, a control framework that leverages an empirical/analytical system model to predict the system's behavior over a short time horizon and provide optimal control inputs accordingly. The MPC solver has been implemented on board, enabling the brainbot to take autonomous actions based on its current state. Considering the computational constraints of the brainbot, TinyMPC~\cite{Nguyen2024} has been chosen for solving the control problem due to its well-balanced computational speed and memory footprint, and adaptability to small embedded systems. Moreover, its operation on the same class of microprocessors (ARM$^\text{\textregistered}$ Cortex series) as used in the brainbot has been demonstrated a priori~\cite{Nguyen2024,ccintacs2026robust}.  

To be able to solve the control problem, TinyMPC requires the state of the brainbot $s=[x,y,v_x,v_y]^\top$, corresponding to its current position $(x,y)$, and velocities of its geometric center in the laboratory frame $(v_x, v_y)$, respectively. The orientation of the bot $\varphi$, although not a part of the state variable, is also required for a subsequent step of the control workflow. Experimentally, the state is obtained through a positioning system using an optical camera and an Aruco marker attached to the brainbot, yielding real-time tracking of the brainbot's position, orientation, and velocity. These measurements are then transmitted to the brainbot at a frequency of $30$ Hz through a Bluetooth protocol, initializing the finite horizon quadratic problem (QP) that is solved by TinyMPC. The full pipeline of the control process is shown in Fig.~\ref{fig:fig8} (a).

The solver predicts the evolution of the state of the bot over a horizon of predefined length $N$ via the linear system model explained later, searching for the sequence of control inputs $\{u^{(j)}\,  \mid \, j=k, \ldots k+N-1\}$ ($k$ denoting the current time step), that minimizes the cost function, while constrained to the imposed state and control input bounds. Only the first control input of the sequence, $u^{(k)}$, is used for the state update, and the remaining sequence is discarded as the uncertainty associated with the prediction accumulates along the horizon. Eq.~(\ref{eq:cost_func_minimize}) delineates the quadratic problem involving the cost function inspired by~\cite{Nguyen2024},
\begin{equation}
\begin{aligned}
\min_{\{u^{(j)}\}_{j=k}^{k+N-1}} \quad & \sum_{j=k}^{k+N-1} \left( \Delta s^{(j)\top} Q\,\Delta s^{(j)} + u^{(j)\top} R\,u^{(j)} \right) \\
 & + \Delta s^{(k+N)\top} P\,\Delta s^{(k+N)}, \\
\text{subject to} \quad & s^{(j+1)} = A_\mathrm{d}\, s^{(j)} + B_\mathrm{d}\, u^{(j)},\\
& u_{\min} \leq u^{(j)} \leq u_{\max},\\
& s_{\min} \leq s^{(j)} \leq s_{\max},\\
\end{aligned}
\label{eq:cost_func_minimize}
\end{equation}
where $s^{(k)} = [x^{(k)},\, y^{(k)},\, v_{x}^{(k)},\, v_{y}^{(k)}]^\top$ and $u^{(k)} = [a_{x}^{(k)},\, a_{y}^{(k)}]^\top$ are the state and control input at the $k^{\text{th}}$ time step, $N$ is the prediction horizon, and $\Delta s^{(k+j)} = s^{(k+j)} - s^{(k+j)}_{\text{ref}}$ is the deviation of the current/predicted state from that of the reference state.

The cost function consists of three terms, from left to right: the tracking cost, the control effort, and the terminal cost, with $Q$, $P$, and $R$ being positive semi-definite cost matrices. The matrices $Q$ and $R$ are diagonal matrices with the values representing the weights of the respective costs, while the matrix $P$ is a symmetric matrix generated by TinyMPC as a solution to the Riccati equation~\cite{underactuated}. The weights for the costs are chosen heuristically from a preliminary \textit{in silico} implementation of the TinyMPC pipeline, replacing the external positioning setup with a simulation-based state measurement of the virtual bot. These weights are then applied to the experimental implementation (see Sec.~IV of Supplementary Information for the initialization of the cost matrices~\cite{supplementary}). The tracking cost penalizes the deviation of the predicted state from the reference state, while the control effort penalizes the large control inputs for state update. Together, these costs account for the stage cost and are accumulated over the prediction horizon $\{k, \ldots k+N-1\}$, where $k$ is the current time step. Terminal cost, on the other hand, represents the state deviation at the terminal step $k+N$, and is essential to ensure numerical stability of the optimization scheme and to prevent undesirable state update beyond the prediction horizon. 
$u_{\min}$, $u_{\max}$, $s_{\min}$ and $s_{\max}$ are the input and state bounds.

TinyMPC owes its efficiency in part to the use of linearized models, namely through the linear double-integrator, with a fixed prediction time step $\Delta t$,
\begin{equation}
s^{(k+1)} = A_\mathrm{d}\, s^{(k)} + B_\mathrm{d} \,u^{(k)},
\end{equation}
where the discretized system matrices with prediction time step $\Delta t$ are
\[
A_\mathrm{d} =
\begin{bmatrix}
1 & 0 & \Delta t & 0 \\
0 & 1 & 0 & \Delta t \\
0 & 0 & 1 & 0 \\
0 & 0 & 0 & 1
\end{bmatrix},
\qquad
B_\mathrm{d} =
\begin{bmatrix}
\frac{1}{2}\Delta t^2 & 0 \\
0 & \frac{1}{2}\Delta t^2 \\
\Delta t & 0 \\
0 & \Delta t
\end{bmatrix}.
\]
The system model assumes the bot to be a point mass located at its geometric center. To ensure consistency with the physical brainbot, the kinematics laid out in Sec.~\ref{sec:kinematics} are reintroduced into the control pipeline in the feedback linearization layer. The virtual control input $[a_x^{(k)}, a_y^{(k)}]^\top$ provided by TinyMPC after solving the minimization problem represents the desired accelerations of the geometric center in the laboratory frame and is used to obtain updated virtual velocities $v_x$ and $v_y$. These virtual velocities, along with the orientation $\varphi$ transmitted by the positioning setup, are then mapped to a desired absolute value of the velocity $v_{\text{command}}$ of the geometric center of the bot and a desired angular velocity $\omega_{\text{command}}$ about the center of rotation of the bot, through Eqs.~\eqref{eq:vfull} and \eqref{eq:drdt}. These values serve as a command to the bot and are the velocity values it is supposed to execute.  

The subsequent step in the control pipeline involves the conversion of these virtual desired values into motion of the physical brainbot by the modulation of the power inputs to the left and right motors $\left(V_L, \, V_R\right)$. This conversion is carried out using lookup tables that define the motor inputs as functions of $v_{\text{command}}$ and $\omega_{\text{command}}$. The lookup tables are pre-built from empirical calibration data and stored in the bot's memory. The calibration involves measuring the $v_{\text{command}}$ and $\omega_{\text{command}}$ over different combinations of motor inputs, followed by fitting the data to a non-linear regression model, and then inverting the model using a least-squares method.




The performance of the embedded model predictive control on the brainbot is demonstrated through a path-following exercise, using a lemniscate as the reference trajectory for the bot. The lemniscate is parameterized as $x=A\sin(2\pi t/T)$, $y=B\sin(2\pi t/T)\cos(2\pi t/T)$, where $T$ allows control over the period of completion of one lemniscate cycle. The geometric parameters of the lemniscate are set to $A=15$ cm and $B=30$ cm based on the field of view of the camera, while the time period is set to $T=28$ s to avoid very low translational and angular velocities. Using the positioning system, the position $\left(x(t),y(t)\right)$ of the bot is recorded, allowing us to trace the path executed by the physical bot as shown in Fig.~\ref{fig:fig8} (b), over approximately sixteen consecutive instances of the reference trajectory. In dashed lines we show the mean trajectory of the bot, which is determined by taking advantage of the periodicity of $x(t)$ and $y(t)$, and averaging their periods. 

The results show that the bot is able to systematically reproduce a lemniscate trajectory with a small error, consistent between different realizations of the trajectory and in agreement with previous implementations of TinyMPC~\cite{Nguyen2024}. Besides this generally good performance, an overall asymmetry of the trajectory is also observed, likely due to the inherent mechanical imperfections of the bot in turning left or right, as well as to the approximation introduced by the linearization of control. Despite the slight deviations, this experiment serves as a clear validation of the ability of the brainbot to solve a control problem onboard, utilizing a kinematics-informed MPC approach, while also providing a reliable experimental platform for a small-scale intelligent active particle capable of autonomous decision-making based on predictive control. To the best of our knowledge, this is the first realization of a bristlebot using onboard model predictive control for navigation of its environment.


\section{Discussion and Conclusion}

We have presented a kinematic model that clearly decomposes the complex trajectories observed in experimental realizations of brainbot motion into basic modes of motion. Thereby we are able to decouple the linear, spinning, and orbital components. The model is agnostic of the dependence of the linear and angular velocities of the brainbot on the driving parameters, 
which is necessary because the motion of the brainbot exhibits a complex dependence on a variety of factors such as the leg shape and angle, the motor voltage, and surface irregularities. Nevertheless, once the velocities are prescribed, the forward kinematic framework reproduces all the experimentally-observed trajectories. The model also uncovers at least one mode of motion, that of a helical closed trajectory (shown in Fig.~\ref{fig:sim1} (f)) that has so far not been observed in the experiments.

We have integrated this physical understanding into a forward kinematic framework, which was trained on experimental data using an offline parametric identification approach to create a digital representation of a brainbot. 
Apart from reproducing experimental trajectories, we have created synthetic trajectories, which differ from the experiments in controlled ways yet mimic their desired properties. For instance, we have demonstrated how we can (i) pick random values of the angular velocity $\omega$ from its experimentally-obtained distribution, or (ii) incorporate different levels of noise in $\omega$ by including an increasing number of modes from the Fourier decomposition of the experimental $\omega$. This allows us to generate simulation trajectories that exhibit distributions of the curvature parameter $\eta$ that are similar to those found from the experiments. This process effectively yields a functional digital replica of the brainbot.

These digital twins are important for providing data both on an infinite domain and for long time periods, overcoming experimental limitations. In previous work, we have shown that brainbots can be programmed to execute diffusive motion on long time scales, while on short time scales they exhibit ballistic motion~\cite{Noirhomme2025, Novkoski2025}. The framework constructed here can be used to find more efficient algorithms of motion that lead to diffusion on a given time scale, without the need for a large number of experiments. It also allows an extension of the bot behavior to more complicated desired regimes, such as super-diffusive and sub-diffusive motion. Additionally, access to a large ensemble of numerical trajectories can prove useful in training machine learning models. Such approaches can learn from the generally stochastic statistics of brainbot trajectories to train neural networks to achieve complex desired trajectories, while incorporating specific interactions between multiple bots, which may possibly reveal novel dynamical regimes. We plan to implement such machine learning methods in future work.

Relying on a kinematic framework has enabled us to demonstrate the first-ever on-board predictive control on a bristlebot to efficiently track a reference trajectory. Although attempts have been made in the past to introduce basic control onto bristlebots~\cite{Mousavi2025}, these rely on complex physical modeling of the bristles. Under our framework, a model predictive control approach that runs directly on the brainbot~\cite{Nguyen2024}~\cite{mahajan2024code} has been leveraged against the kinematic model to execute the tracking of a desired trajectory. The framework is computationally light, making it suitable for onboard applications on microprocessors with low compute capability. While certain tradeoffs exist, most notably the lack of precision due to the linearization of the underlying kinematic problem and the asymmetry of the bot motion, these could be addressed by further improvements in processing power and the use of more
powerful solvers such as OSQP~\cite{osqp} or more efficient optimal control toolchains such as Acados~\cite{verschueren2022acados}, which relies on fast QP solvers. Nevertheless, the demonstrated proof-of-concept showcases onboard control as a promising route towards the use of brainbots as experimental realizations of decentralized intelligent active particles.



The brainbot is a convenient experimental system that is easy to fabricate and program, and versatile in the range of motion it can undergo, either autonomously or through hard-coding. Here, we have demonstrated a robust theoretical and experimental control of its motion.

\acknowledgments
N.V.\ thanks the Fondation Francqui for support. F.N.\ thanks the Alexander von Humboldt Foundation for a postdoctoral fellowship. M.N.\ thanks the Belgian Federal Science Policy Office (BELSPO) for financial support within the PRODEX Programme of the European Space Agency (ESA), Contract No.\ 4000103267.

\bigskip

\bibliography{references}

\begin{thebibliography}{33}%
\makeatletter
\providecommand \@ifxundefined [1]{%
 \@ifx{#1\undefined}
}%
\providecommand \@ifnum [1]{%
 \ifnum #1\expandafter \@firstoftwo
 \else \expandafter \@secondoftwo
 \fi
}%
\providecommand \@ifx [1]{%
 \ifx #1\expandafter \@firstoftwo
 \else \expandafter \@secondoftwo
 \fi
}%
\providecommand \natexlab [1]{#1}%
\providecommand \enquote  [1]{``#1''}%
\providecommand \bibnamefont  [1]{#1}%
\providecommand \bibfnamefont [1]{#1}%
\providecommand \citenamefont [1]{#1}%
\providecommand \href@noop [0]{\@secondoftwo}%
\providecommand \href [0]{\begingroup \@sanitize@url \@href}%
\providecommand \@href[1]{\@@startlink{#1}\@@href}%
\providecommand \@@href[1]{\endgroup#1\@@endlink}%
\providecommand \@sanitize@url [0]{\catcode `\\12\catcode `\$12\catcode `\&12\catcode `\#12\catcode `\^12\catcode `\_12\catcode `\%12\relax}%
\providecommand \@@startlink[1]{}%
\providecommand \@@endlink[0]{}%
\providecommand \url  [0]{\begingroup\@sanitize@url \@url }%
\providecommand \@url [1]{\endgroup\@href {#1}{\urlprefix }}%
\providecommand \urlprefix  [0]{URL }%
\providecommand \Eprint [0]{\href }%
\providecommand \doibase [0]{https://doi.org/}%
\providecommand \selectlanguage [0]{\@gobble}%
\providecommand \bibinfo  [0]{\@secondoftwo}%
\providecommand \bibfield  [0]{\@secondoftwo}%
\providecommand \translation [1]{[#1]}%
\providecommand \BibitemOpen [0]{}%
\providecommand \bibitemStop [0]{}%
\providecommand \bibitemNoStop [0]{.\EOS\space}%
\providecommand \EOS [0]{\spacefactor3000\relax}%
\providecommand \BibitemShut  [1]{\csname bibitem#1\endcsname}%
\let\auto@bib@innerbib\@empty
\bibitem [{\citenamefont {Bechinger}\ \emph {et~al.}(2016)\citenamefont {Bechinger}, \citenamefont {Di~Leonardo}, \citenamefont {L\"owen}, \citenamefont {Reichhardt}, \citenamefont {Volpe},\ and\ \citenamefont {Volpe}}]{Bechinger2016}%
  \BibitemOpen
  \bibfield  {author} {\bibinfo {author} {\bibfnamefont {C.}~\bibnamefont {Bechinger}}, \bibinfo {author} {\bibfnamefont {R.}~\bibnamefont {Di~Leonardo}}, \bibinfo {author} {\bibfnamefont {H.}~\bibnamefont {L\"owen}}, \bibinfo {author} {\bibfnamefont {C.}~\bibnamefont {Reichhardt}}, \bibinfo {author} {\bibfnamefont {G.}~\bibnamefont {Volpe}},\ and\ \bibinfo {author} {\bibfnamefont {G.}~\bibnamefont {Volpe}},\ }\bibfield  {title} {\bibinfo {title} {Active particles in complex and crowded environments},\ }\href {https://doi.org/10.1103/RevModPhys.88.045006} {\bibfield  {journal} {\bibinfo  {journal} {Rev. Mod. Phys.}\ }\textbf {\bibinfo {volume} {88}},\ \bibinfo {pages} {045006} (\bibinfo {year} {2016})}\BibitemShut {NoStop}%
\bibitem [{\citenamefont {Liu}\ \emph {et~al.}(2021)\citenamefont {Liu}, \citenamefont {Shi}, \citenamefont {Zhao}, \citenamefont {Chaté}, \citenamefont {qing Shi},\ and\ \citenamefont {Zhang}}]{Liu2021}%
  \BibitemOpen
  \bibfield  {author} {\bibinfo {author} {\bibfnamefont {Z.~T.}\ \bibnamefont {Liu}}, \bibinfo {author} {\bibfnamefont {Y.}~\bibnamefont {Shi}}, \bibinfo {author} {\bibfnamefont {Y.}~\bibnamefont {Zhao}}, \bibinfo {author} {\bibfnamefont {H.}~\bibnamefont {Chaté}}, \bibinfo {author} {\bibfnamefont {X.}~\bibnamefont {qing Shi}},\ and\ \bibinfo {author} {\bibfnamefont {T.~H.}\ \bibnamefont {Zhang}},\ }\bibfield  {title} {\bibinfo {title} {Activity waves and freestanding vortices in populations of subcritical quincke rollers},\ }\href {https://doi.org/10.1073/pnas.2104724118} {\bibfield  {journal} {\bibinfo  {journal} {Proc. Natl. Acad. Sci. U.S.A.}\ }\textbf {\bibinfo {volume} {118}},\ \bibinfo {pages} {e2104724118} (\bibinfo {year} {2021})}\BibitemShut {NoStop}%
\bibitem [{\citenamefont {Garza}\ \emph {et~al.}(2023)\citenamefont {Garza}, \citenamefont {Kyriakopoulos}, \citenamefont {Cenev}, \citenamefont {Rigoni},\ and\ \citenamefont {Timonen}}]{Garza2023}%
  \BibitemOpen
  \bibfield  {author} {\bibinfo {author} {\bibfnamefont {R.~R.}\ \bibnamefont {Garza}}, \bibinfo {author} {\bibfnamefont {N.}~\bibnamefont {Kyriakopoulos}}, \bibinfo {author} {\bibfnamefont {Z.~M.}\ \bibnamefont {Cenev}}, \bibinfo {author} {\bibfnamefont {C.}~\bibnamefont {Rigoni}},\ and\ \bibinfo {author} {\bibfnamefont {J.~V.~I.}\ \bibnamefont {Timonen}},\ }\bibfield  {title} {\bibinfo {title} {Magnetic quincke rollers with tunable single-particle dynamics and collective states},\ }\href {https://doi.org/10.1126/sciadv.adh2522} {\bibfield  {journal} {\bibinfo  {journal} {Sci. Adv.}\ }\textbf {\bibinfo {volume} {9}},\ \bibinfo {pages} {eadh2522} (\bibinfo {year} {2023})}\BibitemShut {NoStop}%
\bibitem [{\citenamefont {Narayan}\ \emph {et~al.}(2007)\citenamefont {Narayan}, \citenamefont {Ramaswamy},\ and\ \citenamefont {Menon}}]{Narayan2007}%
  \BibitemOpen
  \bibfield  {author} {\bibinfo {author} {\bibfnamefont {V.}~\bibnamefont {Narayan}}, \bibinfo {author} {\bibfnamefont {S.}~\bibnamefont {Ramaswamy}},\ and\ \bibinfo {author} {\bibfnamefont {N.}~\bibnamefont {Menon}},\ }\bibfield  {title} {\bibinfo {title} {Long-lived giant number fluctuations in a swarming granular nematic},\ }\href {https://doi.org/10.1126/science.1140414} {\bibfield  {journal} {\bibinfo  {journal} {Science}\ }\textbf {\bibinfo {volume} {317}},\ \bibinfo {pages} {105} (\bibinfo {year} {2007})}\BibitemShut {NoStop}%
\bibitem [{\citenamefont {Scholz}\ \emph {et~al.}(2018{\natexlab{a}})\citenamefont {Scholz}, \citenamefont {Engel},\ and\ \citenamefont {P{\"o}schel}}]{Scholz2018a}%
  \BibitemOpen
  \bibfield  {author} {\bibinfo {author} {\bibfnamefont {C.}~\bibnamefont {Scholz}}, \bibinfo {author} {\bibfnamefont {M.}~\bibnamefont {Engel}},\ and\ \bibinfo {author} {\bibfnamefont {T.}~\bibnamefont {P{\"o}schel}},\ }\bibfield  {title} {\bibinfo {title} {Rotating robots move collectively and self-organize},\ }\href@noop {} {\bibfield  {journal} {\bibinfo  {journal} {Nat. Commun.}\ }\textbf {\bibinfo {volume} {9}},\ \bibinfo {pages} {931} (\bibinfo {year} {2018}{\natexlab{a}})}\BibitemShut {NoStop}%
\bibitem [{\citenamefont {Scholz}\ \emph {et~al.}(2018{\natexlab{b}})\citenamefont {Scholz}, \citenamefont {Jahanshahi}, \citenamefont {Ldov},\ and\ \citenamefont {L{\"o}wen}}]{Scholz2018b}%
  \BibitemOpen
  \bibfield  {author} {\bibinfo {author} {\bibfnamefont {C.}~\bibnamefont {Scholz}}, \bibinfo {author} {\bibfnamefont {S.}~\bibnamefont {Jahanshahi}}, \bibinfo {author} {\bibfnamefont {A.}~\bibnamefont {Ldov}},\ and\ \bibinfo {author} {\bibfnamefont {H.}~\bibnamefont {L{\"o}wen}},\ }\bibfield  {title} {\bibinfo {title} {Inertial delay of self-propelled particles},\ }\href {https://doi.org/10.1038/s41467-018-07596-x} {\bibfield  {journal} {\bibinfo  {journal} {Nat. Commun.}\ }\textbf {\bibinfo {volume} {9}},\ \bibinfo {pages} {5156} (\bibinfo {year} {2018}{\natexlab{b}})}\BibitemShut {NoStop}%
\bibitem [{\citenamefont {Caprini}\ \emph {et~al.}(2024{\natexlab{a}})\citenamefont {Caprini}, \citenamefont {Breoni}, \citenamefont {Ldov}, \citenamefont {Scholz},\ and\ \citenamefont {L{\"o}wen}}]{Caprini2024}%
  \BibitemOpen
  \bibfield  {author} {\bibinfo {author} {\bibfnamefont {L.}~\bibnamefont {Caprini}}, \bibinfo {author} {\bibfnamefont {D.}~\bibnamefont {Breoni}}, \bibinfo {author} {\bibfnamefont {A.}~\bibnamefont {Ldov}}, \bibinfo {author} {\bibfnamefont {C.}~\bibnamefont {Scholz}},\ and\ \bibinfo {author} {\bibfnamefont {H.}~\bibnamefont {L{\"o}wen}},\ }\bibfield  {title} {\bibinfo {title} {Dynamical clustering and wetting phenomena in inertial active matter},\ }\href {https://doi.org/10.1038/s42005-024-01835-y} {\bibfield  {journal} {\bibinfo  {journal} {Commun. Phys.}\ }\textbf {\bibinfo {volume} {7}},\ \bibinfo {pages} {343} (\bibinfo {year} {2024}{\natexlab{a}})}\BibitemShut {NoStop}%
\bibitem [{\citenamefont {Chen}\ and\ \citenamefont {Zhang}(2024)}]{Chen2024}%
  \BibitemOpen
  \bibfield  {author} {\bibinfo {author} {\bibfnamefont {Y.}~\bibnamefont {Chen}}\ and\ \bibinfo {author} {\bibfnamefont {J.}~\bibnamefont {Zhang}},\ }\bibfield  {title} {\bibinfo {title} {Anomalous flocking in nonpolar granular brownian vibrators},\ }\href {https://doi.org/10.1038/s41467-024-50479-7} {\bibfield  {journal} {\bibinfo  {journal} {Nat. Commun.}\ }\textbf {\bibinfo {volume} {15}},\ \bibinfo {pages} {6032} (\bibinfo {year} {2024})}\BibitemShut {NoStop}%
\bibitem [{\citenamefont {Caprini}\ \emph {et~al.}(2024{\natexlab{b}})\citenamefont {Caprini}, \citenamefont {Liebchen},\ and\ \citenamefont {L{\"o}wen}}]{Caprini2024b}%
  \BibitemOpen
  \bibfield  {author} {\bibinfo {author} {\bibfnamefont {L.}~\bibnamefont {Caprini}}, \bibinfo {author} {\bibfnamefont {B.}~\bibnamefont {Liebchen}},\ and\ \bibinfo {author} {\bibfnamefont {H.}~\bibnamefont {L{\"o}wen}},\ }\bibfield  {title} {\bibinfo {title} {Self-reverting vortices in chiral active matter},\ }\href {https://doi.org/10.1038/s42005-024-01637-2} {\bibfield  {journal} {\bibinfo  {journal} {Commun. Phys.}\ }\textbf {\bibinfo {volume} {7}},\ \bibinfo {pages} {153} (\bibinfo {year} {2024}{\natexlab{b}})}\BibitemShut {NoStop}%
\bibitem [{\citenamefont {Dauchot}\ and\ \citenamefont {D\'emery}(2019)}]{Dauchot2019}%
  \BibitemOpen
  \bibfield  {author} {\bibinfo {author} {\bibfnamefont {O.}~\bibnamefont {Dauchot}}\ and\ \bibinfo {author} {\bibfnamefont {V.}~\bibnamefont {D\'emery}},\ }\bibfield  {title} {\bibinfo {title} {Dynamics of a self-propelled particle in a harmonic trap},\ }\href {https://doi.org/10.1103/PhysRevLett.122.068002} {\bibfield  {journal} {\bibinfo  {journal} {Phys. Rev. Lett.}\ }\textbf {\bibinfo {volume} {122}},\ \bibinfo {pages} {068002} (\bibinfo {year} {2019})}\BibitemShut {NoStop}%
\bibitem [{\citenamefont {Baconnier}\ \emph {et~al.}(2022)\citenamefont {Baconnier}, \citenamefont {Shohat}, \citenamefont {L{\'o}pez}, \citenamefont {Coulais}, \citenamefont {D{\'e}mery}, \citenamefont {D{\"u}ring},\ and\ \citenamefont {Dauchot}}]{Baconnier2022}%
  \BibitemOpen
  \bibfield  {author} {\bibinfo {author} {\bibfnamefont {P.}~\bibnamefont {Baconnier}}, \bibinfo {author} {\bibfnamefont {D.}~\bibnamefont {Shohat}}, \bibinfo {author} {\bibfnamefont {C.~H.}\ \bibnamefont {L{\'o}pez}}, \bibinfo {author} {\bibfnamefont {C.}~\bibnamefont {Coulais}}, \bibinfo {author} {\bibfnamefont {V.}~\bibnamefont {D{\'e}mery}}, \bibinfo {author} {\bibfnamefont {G.}~\bibnamefont {D{\"u}ring}},\ and\ \bibinfo {author} {\bibfnamefont {O.}~\bibnamefont {Dauchot}},\ }\bibfield  {title} {\bibinfo {title} {Selective and collective actuation in active solids},\ }\href {https://doi.org/10.1038/s41567-022-01704-x} {\bibfield  {journal} {\bibinfo  {journal} {Nat. Phys.}\ }\textbf {\bibinfo {volume} {18}},\ \bibinfo {pages} {1234} (\bibinfo {year} {2022})}\BibitemShut {NoStop}%
\bibitem [{\citenamefont {Chor}\ \emph {et~al.}(2023)\citenamefont {Chor}, \citenamefont {Sohachi}, \citenamefont {Goerlich}, \citenamefont {Rosen}, \citenamefont {Rahav},\ and\ \citenamefont {Roichman}}]{Chor2023}%
  \BibitemOpen
  \bibfield  {author} {\bibinfo {author} {\bibfnamefont {O.}~\bibnamefont {Chor}}, \bibinfo {author} {\bibfnamefont {A.}~\bibnamefont {Sohachi}}, \bibinfo {author} {\bibfnamefont {R.}~\bibnamefont {Goerlich}}, \bibinfo {author} {\bibfnamefont {E.}~\bibnamefont {Rosen}}, \bibinfo {author} {\bibfnamefont {S.}~\bibnamefont {Rahav}},\ and\ \bibinfo {author} {\bibfnamefont {Y.}~\bibnamefont {Roichman}},\ }\bibfield  {title} {\bibinfo {title} {Many-body szil\'ard engine with giant number fluctuations},\ }\href {https://doi.org/10.1103/PhysRevResearch.5.043193} {\bibfield  {journal} {\bibinfo  {journal} {Phys. Rev. Res.}\ }\textbf {\bibinfo {volume} {5}},\ \bibinfo {pages} {043193} (\bibinfo {year} {2023})}\BibitemShut {NoStop}%
\bibitem [{\citenamefont {Giomi}\ \emph {et~al.}(2013)\citenamefont {Giomi}, \citenamefont {Hawley-Weld},\ and\ \citenamefont {Mahadevan}}]{Giomi2013}%
  \BibitemOpen
  \bibfield  {author} {\bibinfo {author} {\bibfnamefont {L.}~\bibnamefont {Giomi}}, \bibinfo {author} {\bibfnamefont {N.}~\bibnamefont {Hawley-Weld}},\ and\ \bibinfo {author} {\bibfnamefont {L.}~\bibnamefont {Mahadevan}},\ }\bibfield  {title} {\bibinfo {title} {Swarming, swirling and stasis in sequestered bristle-bots},\ }\href {https://doi.org/10.1098/rspa.2012.0637} {\bibfield  {journal} {\bibinfo  {journal} {Proc. R. Soc. A}\ }\textbf {\bibinfo {volume} {469}},\ \bibinfo {pages} {20120637} (\bibinfo {year} {2013})}\BibitemShut {NoStop}%
\bibitem [{\citenamefont {Zheng}\ \emph {et~al.}(2023)\citenamefont {Zheng}, \citenamefont {Brandenbourger}, \citenamefont {Robinet}, \citenamefont {Schall}, \citenamefont {Lerner},\ and\ \citenamefont {Coulais}}]{Zheng2023}%
  \BibitemOpen
  \bibfield  {author} {\bibinfo {author} {\bibfnamefont {E.}~\bibnamefont {Zheng}}, \bibinfo {author} {\bibfnamefont {M.}~\bibnamefont {Brandenbourger}}, \bibinfo {author} {\bibfnamefont {L.}~\bibnamefont {Robinet}}, \bibinfo {author} {\bibfnamefont {P.}~\bibnamefont {Schall}}, \bibinfo {author} {\bibfnamefont {E.}~\bibnamefont {Lerner}},\ and\ \bibinfo {author} {\bibfnamefont {C.}~\bibnamefont {Coulais}},\ }\bibfield  {title} {\bibinfo {title} {Self-oscillation and synchronization transitions in elastoactive structures},\ }\href {https://doi.org/10.1103/PhysRevLett.130.178202} {\bibfield  {journal} {\bibinfo  {journal} {Phys. Rev. Lett.}\ }\textbf {\bibinfo {volume} {130}},\ \bibinfo {pages} {178202} (\bibinfo {year} {2023})}\BibitemShut {NoStop}%
\bibitem [{\citenamefont {Altshuler}\ \emph {et~al.}(2024)\citenamefont {Altshuler}, \citenamefont {Bonomo}, \citenamefont {Gorohovsky}, \citenamefont {Marchini}, \citenamefont {Rosen}, \citenamefont {Tal-Friedman}, \citenamefont {Reuveni},\ and\ \citenamefont {Roichman}}]{Altshuler2024}%
  \BibitemOpen
  \bibfield  {author} {\bibinfo {author} {\bibfnamefont {A.}~\bibnamefont {Altshuler}}, \bibinfo {author} {\bibfnamefont {O.~L.}\ \bibnamefont {Bonomo}}, \bibinfo {author} {\bibfnamefont {N.}~\bibnamefont {Gorohovsky}}, \bibinfo {author} {\bibfnamefont {S.}~\bibnamefont {Marchini}}, \bibinfo {author} {\bibfnamefont {E.}~\bibnamefont {Rosen}}, \bibinfo {author} {\bibfnamefont {O.}~\bibnamefont {Tal-Friedman}}, \bibinfo {author} {\bibfnamefont {S.}~\bibnamefont {Reuveni}},\ and\ \bibinfo {author} {\bibfnamefont {Y.}~\bibnamefont {Roichman}},\ }\bibfield  {title} {\bibinfo {title} {Environmental memory facilitates search with home returns},\ }\href {https://doi.org/10.1103/PhysRevResearch.6.023255} {\bibfield  {journal} {\bibinfo  {journal} {Phys. Rev. Res.}\ }\textbf {\bibinfo {volume} {6}},\ \bibinfo {pages} {023255} (\bibinfo {year} {2024})}\BibitemShut {NoStop}%
\bibitem [{\citenamefont {Noirhomme}\ \emph {et~al.}(2025)\citenamefont {Noirhomme}, \citenamefont {Mammadli}, \citenamefont {Vanesse}, \citenamefont {Pande}, \citenamefont {Smith},\ and\ \citenamefont {Vandewalle}}]{Noirhomme2025}%
  \BibitemOpen
  \bibfield  {author} {\bibinfo {author} {\bibfnamefont {M.}~\bibnamefont {Noirhomme}}, \bibinfo {author} {\bibfnamefont {I.}~\bibnamefont {Mammadli}}, \bibinfo {author} {\bibfnamefont {N.}~\bibnamefont {Vanesse}}, \bibinfo {author} {\bibfnamefont {J.}~\bibnamefont {Pande}}, \bibinfo {author} {\bibfnamefont {A.-S.}\ \bibnamefont {Smith}},\ and\ \bibinfo {author} {\bibfnamefont {N.}~\bibnamefont {Vandewalle}},\ }\bibfield  {title} {\bibinfo {title} {Brainbots as smart autonomous active particles with programmable motion},\ }\href {https://doi.org/10.1103/pv3d-7s54} {\bibfield  {journal} {\bibinfo  {journal} {Phys. Rev. Appl.}\ }\textbf {\bibinfo {volume} {23}},\ \bibinfo {pages} {064008} (\bibinfo {year} {2025})}\BibitemShut {NoStop}%
\bibitem [{\citenamefont {Novkoski}\ \emph {et~al.}(2026)\citenamefont {Novkoski}, \citenamefont {M{\'e}lard}, \citenamefont {Delens}, \citenamefont {W{\'e}ry}, \citenamefont {Noirhomme}, \citenamefont {Pande}, \citenamefont {Maier}, \citenamefont {Smith},\ and\ \citenamefont {Vandewalle}}]{Novkoski2025}%
  \BibitemOpen
  \bibfield  {author} {\bibinfo {author} {\bibfnamefont {F.}~\bibnamefont {Novkoski}}, \bibinfo {author} {\bibfnamefont {M.}~\bibnamefont {M{\'e}lard}}, \bibinfo {author} {\bibfnamefont {M.}~\bibnamefont {Delens}}, \bibinfo {author} {\bibfnamefont {F.}~\bibnamefont {W{\'e}ry}}, \bibinfo {author} {\bibfnamefont {M.}~\bibnamefont {Noirhomme}}, \bibinfo {author} {\bibfnamefont {J.}~\bibnamefont {Pande}}, \bibinfo {author} {\bibfnamefont {A.}~\bibnamefont {Maier}}, \bibinfo {author} {\bibfnamefont {A.-S.}\ \bibnamefont {Smith}},\ and\ \bibinfo {author} {\bibfnamefont {N.}~\bibnamefont {Vandewalle}},\ }\bibfield  {title} {\bibinfo {title} {Graspion: An open-source, programmable brainbot for active matter research},\ }\href {https://doi.org/10.1063/5.0303152} {\bibfield  {journal} {\bibinfo  {journal} {Review of Scientific Instruments}\ }\textbf {\bibinfo {volume} {97}} (\bibinfo {year} {2026})}\BibitemShut {NoStop}%
\bibitem [{\citenamefont {Becker}\ \emph {et~al.}(2014)\citenamefont {Becker}, \citenamefont {Boerner}, \citenamefont {Lysenko}, \citenamefont {Zeidis},\ and\ \citenamefont {Zimmermann}}]{Becker2014}%
  \BibitemOpen
  \bibfield  {author} {\bibinfo {author} {\bibfnamefont {F.}~\bibnamefont {Becker}}, \bibinfo {author} {\bibfnamefont {S.}~\bibnamefont {Boerner}}, \bibinfo {author} {\bibfnamefont {V.}~\bibnamefont {Lysenko}}, \bibinfo {author} {\bibfnamefont {I.}~\bibnamefont {Zeidis}},\ and\ \bibinfo {author} {\bibfnamefont {K.}~\bibnamefont {Zimmermann}},\ }\bibfield  {title} {\bibinfo {title} {On the mechanics of bristle-bots - modeling, simulation and experiments},\ }in\ \href@noop {} {\emph {\bibinfo {booktitle} {ISR/Robotik 2014; 41st International Symposium on Robotics}}}\ (\bibinfo {year} {2014})\ pp.\ \bibinfo {pages} {1--6}\BibitemShut {NoStop}%
\bibitem [{\citenamefont {Majewski}\ \emph {et~al.}(2017)\citenamefont {Majewski}, \citenamefont {Szwedowicz},\ and\ \citenamefont {Majewski}}]{Majewski2017}%
  \BibitemOpen
  \bibfield  {author} {\bibinfo {author} {\bibfnamefont {T.}~\bibnamefont {Majewski}}, \bibinfo {author} {\bibfnamefont {D.}~\bibnamefont {Szwedowicz}},\ and\ \bibinfo {author} {\bibfnamefont {M.}~\bibnamefont {Majewski}},\ }\bibfield  {title} {\bibinfo {title} {Locomotion of a mini bristle robot with inertial excitation},\ }\href {https://doi.org/10.1115/1.4037892} {\bibfield  {journal} {\bibinfo  {journal} {J. Mech. Robot}\ }\textbf {\bibinfo {volume} {9}},\ \bibinfo {pages} {061008} (\bibinfo {year} {2017})}\BibitemShut {NoStop}%
\bibitem [{\citenamefont {L{\"o}wen}\ and\ \citenamefont {Liebchen}(2026)}]{Lowen2026}%
  \BibitemOpen
  \bibfield  {author} {\bibinfo {author} {\bibfnamefont {H.}~\bibnamefont {L{\"o}wen}}\ and\ \bibinfo {author} {\bibfnamefont {B.}~\bibnamefont {Liebchen}},\ }\bibfield  {title} {\bibinfo {title} {Towards intelligent active particles},\ }in\ \href@noop {} {\emph {\bibinfo {booktitle} {Artificial Intelligence and Intelligent Matter: Nanoscience, Soft Matter, Philosophy}}}\ (\bibinfo  {publisher} {Springer},\ \bibinfo {year} {2026})\ pp.\ \bibinfo {pages} {257--271}\BibitemShut {NoStop}%
\bibitem [{\citenamefont {Mahajan}\ and\ \citenamefont {Plancher}(2025)}]{11246255}%
  \BibitemOpen
  \bibfield  {author} {\bibinfo {author} {\bibfnamefont {I.}~\bibnamefont {Mahajan}}\ and\ \bibinfo {author} {\bibfnamefont {B.}~\bibnamefont {Plancher}},\ }\bibfield  {title} {\bibinfo {title} {Robust and efficient embedded convex optimization through first-order adaptive caching},\ }in\ \href {https://doi.org/10.1109/IROS60139.2025.11246255} {\emph {\bibinfo {booktitle} {2025 IEEE/RSJ International Conference on Intelligent Robots and Systems (IROS)}}}\ (\bibinfo {year} {2025})\ pp.\ \bibinfo {pages} {2984--2990}\BibitemShut {NoStop}%
\bibitem [{\citenamefont {Nguyen}\ \emph {et~al.}(2024)\citenamefont {Nguyen}, \citenamefont {Schoedel}, \citenamefont {Alavilli}, \citenamefont {Plancher},\ and\ \citenamefont {Manchester}}]{Nguyen2024}%
  \BibitemOpen
  \bibfield  {author} {\bibinfo {author} {\bibfnamefont {K.}~\bibnamefont {Nguyen}}, \bibinfo {author} {\bibfnamefont {S.}~\bibnamefont {Schoedel}}, \bibinfo {author} {\bibfnamefont {A.}~\bibnamefont {Alavilli}}, \bibinfo {author} {\bibfnamefont {B.}~\bibnamefont {Plancher}},\ and\ \bibinfo {author} {\bibfnamefont {Z.}~\bibnamefont {Manchester}},\ }\bibfield  {title} {\bibinfo {title} {Tinympc: Model-predictive control on resource-constrained microcontrollers},\ }in\ \href@noop {} {\emph {\bibinfo {booktitle} {2024 IEEE International Conference on Robotics and Automation (ICRA)}}}\ (\bibinfo {organization} {IEEE},\ \bibinfo {year} {2024})\ pp.\ \bibinfo {pages} {1--7}\BibitemShut {NoStop}%
\bibitem [{\citenamefont {Wikner}\ \emph {et~al.}(2026)\citenamefont {Wikner}, \citenamefont {Arnström},\ and\ \citenamefont {Axehill}}]{wikner2026realtimefeasibilityhighratempc}%
  \BibitemOpen
  \bibfield  {author} {\bibinfo {author} {\bibfnamefont {J.}~\bibnamefont {Wikner}}, \bibinfo {author} {\bibfnamefont {D.}~\bibnamefont {Arnström}},\ and\ \bibinfo {author} {\bibfnamefont {D.}~\bibnamefont {Axehill}},\ }\href {https://arxiv.org/abs/2603.09342} {\bibinfo {title} {On real-time feasibility of high-rate mpc using an active-set method on nano-quadcopters}} (\bibinfo {year} {2026}),\ \Eprint {https://arxiv.org/abs/2603.09342} {arXiv:2603.09342 [math.OC]} \BibitemShut {NoStop}%
\bibitem [{sup()}]{supplementary}%
  \BibitemOpen
  \href@noop {} {}\bibinfo {note} {See Supplemental Material at [URL will be inserted by publisher] for equivalence of constant velocity in laboratory frame and sinusoidal velocity in body-fixed frame, the numerical integration scheme used in the simulations, comparison between experiments and simulations for different numbers of Fourier modes and the values of the cost matrices used, which includes Refs.~\cite{Crank1947}}\BibitemShut {NoStop}%
\bibitem [{\citenamefont {Savitzky}\ and\ \citenamefont {Golay}(1964)}]{savgol}%
  \BibitemOpen
  \bibfield  {author} {\bibinfo {author} {\bibfnamefont {A.}~\bibnamefont {Savitzky}}\ and\ \bibinfo {author} {\bibfnamefont {M.~J.~E.}\ \bibnamefont {Golay}},\ }\bibfield  {title} {\bibinfo {title} {Smoothing and differentiation of data by simplified least squares procedures.},\ }\href {https://doi.org/10.1021/ac60214a047} {\bibfield  {journal} {\bibinfo  {journal} {Anal. Chem.}\ }\textbf {\bibinfo {volume} {36}},\ \bibinfo {pages} {1627} (\bibinfo {year} {1964})},\ \Eprint {https://arxiv.org/abs/https://doi.org/10.1021/ac60214a047} {https://doi.org/10.1021/ac60214a047} \BibitemShut {NoStop}%
\bibitem [{\citenamefont {Kouvaritakis}\ and\ \citenamefont {Cannon}(2016)}]{kouvaritakis2016model}%
  \BibitemOpen
  \bibfield  {author} {\bibinfo {author} {\bibfnamefont {B.}~\bibnamefont {Kouvaritakis}}\ and\ \bibinfo {author} {\bibfnamefont {M.}~\bibnamefont {Cannon}},\ }\bibfield  {title} {\bibinfo {title} {Model predictive control},\ }\href@noop {} {\bibfield  {journal} {\bibinfo  {journal} {Switzerland: Springer International Publishing}\ }\textbf {\bibinfo {volume} {38}},\ \bibinfo {pages} {7} (\bibinfo {year} {2016})}\BibitemShut {NoStop}%
\bibitem [{\citenamefont {{\c{C}}inta{\c{s}}}\ and\ \citenamefont {{\"O}zyer}(2026)}]{ccintacs2026robust}%
  \BibitemOpen
  \bibfield  {author} {\bibinfo {author} {\bibfnamefont {E.}~\bibnamefont {{\c{C}}inta{\c{s}}}}\ and\ \bibinfo {author} {\bibfnamefont {B.}~\bibnamefont {{\"O}zyer}},\ }\bibfield  {title} {\bibinfo {title} {A robust fault-tolerant control algorithm for gps-denied mini quadrotors using pid-tinympc and visual-inertial odometry},\ }\href@noop {} {\bibfield  {journal} {\bibinfo  {journal} {Control Engineering Practice}\ }\textbf {\bibinfo {volume} {169}},\ \bibinfo {pages} {106779} (\bibinfo {year} {2026})}\BibitemShut {NoStop}%
\bibitem [{\citenamefont {Tedrake}(2023)}]{underactuated}%
  \BibitemOpen
  \bibfield  {author} {\bibinfo {author} {\bibfnamefont {R.}~\bibnamefont {Tedrake}},\ }\href {https://underactuated.csail.mit.edu} {\emph {\bibinfo {title} {Underactuated Robotics}}}\ (\bibinfo {year} {2023})\BibitemShut {NoStop}%
\bibitem [{\citenamefont {Mousavi}\ and\ \citenamefont {Fakhari}(2025)}]{Mousavi2025}%
  \BibitemOpen
  \bibfield  {author} {\bibinfo {author} {\bibfnamefont {S.~M.}\ \bibnamefont {Mousavi}}\ and\ \bibinfo {author} {\bibfnamefont {V.}~\bibnamefont {Fakhari}},\ }\bibfield  {title} {\bibinfo {title} {Design and development of closed-loop controllers for trajectory tracking of a planar vibration-driven robot},\ }\href {https://doi.org/10.1177/10775463241276026} {\bibfield  {journal} {\bibinfo  {journal} {J. Vib. Control}\ }\textbf {\bibinfo {volume} {31}},\ \bibinfo {pages} {3691} (\bibinfo {year} {2025})}\BibitemShut {NoStop}%
\bibitem [{\citenamefont {Mahajan}\ \emph {et~al.}(2024)\citenamefont {Mahajan}, \citenamefont {Nguyen}, \citenamefont {Schoedel}, \citenamefont {Nedumaran}, \citenamefont {Mata}, \citenamefont {Plancher},\ and\ \citenamefont {Manchester}}]{mahajan2024code}%
  \BibitemOpen
  \bibfield  {author} {\bibinfo {author} {\bibfnamefont {I.}~\bibnamefont {Mahajan}}, \bibinfo {author} {\bibfnamefont {K.}~\bibnamefont {Nguyen}}, \bibinfo {author} {\bibfnamefont {S.}~\bibnamefont {Schoedel}}, \bibinfo {author} {\bibfnamefont {E.}~\bibnamefont {Nedumaran}}, \bibinfo {author} {\bibfnamefont {M.}~\bibnamefont {Mata}}, \bibinfo {author} {\bibfnamefont {B.}~\bibnamefont {Plancher}},\ and\ \bibinfo {author} {\bibfnamefont {Z.}~\bibnamefont {Manchester}},\ }\bibfield  {title} {\bibinfo {title} {Code generation and conic constraints for model-predictive control on microcontrollers with conic-tinympc},\ }\href@noop {} {\bibfield  {journal} {\bibinfo  {journal} {arXiv preprint arXiv:2403.18149}\ } (\bibinfo {year} {2024})}\BibitemShut {NoStop}%
\bibitem [{\citenamefont {Stellato}\ \emph {et~al.}(2020)\citenamefont {Stellato}, \citenamefont {Banjac}, \citenamefont {Goulart}, \citenamefont {Bemporad},\ and\ \citenamefont {Boyd}}]{osqp}%
  \BibitemOpen
  \bibfield  {author} {\bibinfo {author} {\bibfnamefont {B.}~\bibnamefont {Stellato}}, \bibinfo {author} {\bibfnamefont {G.}~\bibnamefont {Banjac}}, \bibinfo {author} {\bibfnamefont {P.}~\bibnamefont {Goulart}}, \bibinfo {author} {\bibfnamefont {A.}~\bibnamefont {Bemporad}},\ and\ \bibinfo {author} {\bibfnamefont {S.}~\bibnamefont {Boyd}},\ }\bibfield  {title} {\bibinfo {title} {{OSQP}: an operator splitting solver for quadratic programs},\ }\href {https://doi.org/10.1007/s12532-020-00179-2} {\bibfield  {journal} {\bibinfo  {journal} {Math. Program. Comput.}\ }\textbf {\bibinfo {volume} {12}},\ \bibinfo {pages} {637} (\bibinfo {year} {2020})}\BibitemShut {NoStop}%
\bibitem [{\citenamefont {Verschueren}\ \emph {et~al.}(2022)\citenamefont {Verschueren}, \citenamefont {Frison}, \citenamefont {Kouzoupis}, \citenamefont {Frey}, \citenamefont {Duijkeren}, \citenamefont {Zanelli}, \citenamefont {Novoselnik}, \citenamefont {Albin}, \citenamefont {Quirynen},\ and\ \citenamefont {Diehl}}]{verschueren2022acados}%
  \BibitemOpen
  \bibfield  {author} {\bibinfo {author} {\bibfnamefont {R.}~\bibnamefont {Verschueren}}, \bibinfo {author} {\bibfnamefont {G.}~\bibnamefont {Frison}}, \bibinfo {author} {\bibfnamefont {D.}~\bibnamefont {Kouzoupis}}, \bibinfo {author} {\bibfnamefont {J.}~\bibnamefont {Frey}}, \bibinfo {author} {\bibfnamefont {N.~v.}\ \bibnamefont {Duijkeren}}, \bibinfo {author} {\bibfnamefont {A.}~\bibnamefont {Zanelli}}, \bibinfo {author} {\bibfnamefont {B.}~\bibnamefont {Novoselnik}}, \bibinfo {author} {\bibfnamefont {T.}~\bibnamefont {Albin}}, \bibinfo {author} {\bibfnamefont {R.}~\bibnamefont {Quirynen}},\ and\ \bibinfo {author} {\bibfnamefont {M.}~\bibnamefont {Diehl}},\ }\bibfield  {title} {\bibinfo {title} {acados—a modular open-source framework for fast embedded optimal control},\ }\href {https://doi.org/10.1007/s12532-021-00208-8} {\bibfield  {journal} {\bibinfo  {journal} {Math. Program. Comput.}\ }\textbf {\bibinfo {volume} {14}},\ \bibinfo {pages} {147} (\bibinfo {year} {2022})}\BibitemShut {NoStop}%
\bibitem [{\citenamefont {Crank}\ and\ \citenamefont {Nicolson}(1947)}]{Crank1947}%
  \BibitemOpen
  \bibfield  {author} {\bibinfo {author} {\bibfnamefont {J.}~\bibnamefont {Crank}}\ and\ \bibinfo {author} {\bibfnamefont {P.}~\bibnamefont {Nicolson}},\ }\bibfield  {title} {\bibinfo {title} {A practical method for numerical evaluation of solutions of partial differential equations of the heat-conduction type},\ }\href {https://doi.org/10.1017/S0305004100023197} {\bibfield  {journal} {\bibinfo  {journal} {Math. Proc. Camb. Philos. Soc.}\ }\textbf {\bibinfo {volume} {43}},\ \bibinfo {pages} {50–67} (\bibinfo {year} {1947})}\BibitemShut {NoStop}%
\end{thebibliography}%

\newpage

\setcounter{equation}{0}
\setcounter{figure}{0}
\setcounter{section}{0}

\renewcommand{\theequation}{S\arabic{equation}}
\renewcommand{\thefigure}{S\arabic{figure}}
\renewcommand{\thesection}{S\arabic{section}}

\section*{Supplemental Material for "Physics-informed digital twin and onboard control of a brainbot for intelligent active matter"}

In this Supplemental Material, we demonstrate the equivalence of constant velocity in laboratory frame and sinusoidal velocity in the body-fixed frame in Sec.~\ref{app:A}, we provide the numerical integration scheme that is used for the simulations in Sec.~\ref{app:B}, the comparison of experimental trajectories to simulated ones as a function of the number of Fourier modes used is given in Sec.~\ref{app:C}, and finally we give the numerical values of the cost matrices used in TinyMPC in Sec.~\ref{app:D}.

\section{Equivalence of constant velocity in laboratory frame and sinusoidal velocity in body-fixed frame}\label{app:A}

Suppose $\bm v_\text{c}$ is a constant vector in the laboratory frame, given by
\begin{align}\label{eq:vccomponents}
\bm v_\text{c} = v_{\text{c}_1} \bm{\hat{i}} + v_{\text{c}_2} \bm{\hat{j}},
\end{align}
where $v_{\text{c}_1}$ and $v_{\text{c}_2}$ are constant. Eq.~(\ref{eq:vccomponents}) can be written as
\begin{align}
\bm v_\text{c} = v_0 \left(\cos \delta \bm{\hat{i}} + \sin \delta \bm{\hat{j}}\right),
\end{align}
where 
\begin{align}
v_0 &= \sqrt{v_{\text{c}_1}^2 + v_{\text{c}_2}^2} \ \text{ and } \nonumber\\
\delta &= \cos^{-1}\left(\dfrac{v_{\text{c}_1}}{v_0}\right) = \sin^{-1}\left(\dfrac{v_{\text{c}_2}}{v_0}\right)
\end{align}
are constants. Therefore, we can write
\begin{align}\label{eq:vc2}
\bm v_\text{c} = v_0 \left\{\cos\left[\varphi(0) + \alpha\right] \bm{\hat{i}} + \sin\left[\varphi(0) + \alpha\right] \bm{\hat{i}}\right\},
\end{align}
where $\varphi(0)$ and $\alpha \equiv \delta - \varphi(0)$ are constants.

Since $\varphi(t) = \varphi(0) + \omega t$, we can write Eq.~(\ref{eq:vc2}) as
\begin{align}
\bm v_\text{c} &= v_0 \left[\cos\left(\varphi(t) - \omega t + \alpha\right)\bm{\hat{i}} + \sin\left(\varphi(t) - \omega t + \alpha\right)\bm{\hat{j}}\right]\nonumber\\
            &= v_0 \left[\cos\left(\alpha - \omega t\right)\cos\varphi(t) \bm{\hat{i}} - \sin\left(\alpha - \omega t\right)\sin\varphi(t) \bm{\hat{i}} \right. \nonumber \\
            & \ \ \ \ \ +\left.\sin\left(\alpha - \omega t\right)\cos\varphi(t) \bm{\hat{j}} + \cos\left(\alpha - \omega t\right)\sin\varphi(t) \bm{\hat{j}}\right]\nonumber\\
            &= v_0 \left[\cos\left(\alpha - \omega t\right)\left(\cos\varphi(t)\bm{\hat{i}} + \sin\varphi(t)\bm{\hat{j}}\right) \right.\nonumber\\
            & \ \ \ \ \ \ \left.+\sin\left(\alpha - \omega t\right)\left(-\sin\varphi(t)\bm{\hat{i}} + \cos\varphi(t)\bm{\hat{j}}\right)\right]\nonumber\\
            &= v_0 \left[\cos\left(\alpha - \omega t\right) \bm{\hat{n}_1} + \sin\left(\alpha - \omega t\right) \bm{\hat{n}_2}\right],
\end{align}
thus showing that in the body-fixed frame, $\bm v_\text{c}$ is a sinusoidal vector.\\

\section{Numerical integration scheme used in the simulations}\label{app:B}


A general time-dependent function $\bm v_\text{c}$ can be integrated numerically using the Crank-Nicolson scheme, accurate to the second order [33],
\begin{equation}
    \int\limits_t^{t+\Delta t} {\bm v}_\text{c} \mathrm dt \approx  \bar {\bm v}_\text{c}(t) \Delta t,
\end{equation}
where $\bar {\bm v}_\text{c}(t)$ is the mean velocity in the interval $t$ to $t + \Delta t$, defined as
\begin{equation}
    \bar {\bm v}_\text{c}(t) \equiv \left( \frac{\bm v_\text{c}(t) + \bm v_\text{c}(t+\Delta t)}{2} \right). 
\end{equation}
Similarly, we define the mean value of the angular velocity in the interval $t$ to $t + \Delta t$ as
\begin{equation}
     \bar \omega(t) \equiv \left( \frac{\omega(t) + \omega(t+\Delta t)}{2} \right). 
\end{equation}
We reformulate Eq.~(10) from the main manuscript in terms of the unit vectors $\bm{\hat n_i}$ while keeping $\bm{v_\text{c}}$ general, thereby obtaining
\begin{align}
    \bm r(t+\Delta t) &= \bm r(t) + \bar{\bm v}_\text{c} \Delta t - \rho_1A_1  [\bm {\hat n_1}(t+\Delta t)-\bm {\hat n_1}(t)] \notag \\  & \ \ \ -\rho_2A_2  [\bm {\hat n_2}(t+\Delta t)-\bm {\hat n_2}(t)].
    \label{eq:rfullv2_mat}
\end{align}
Next, we write the unit vectors in terms of a rotation matrix,
\begin{equation}\label{eq:B5}
    \hat {\bm n} (\varphi(t+\Delta t)) = \bm M(\bar \omega \Delta t)\hat {\bm n}(\varphi(t))
\end{equation}
where 
\begin{equation}\label{eq:B6}
\bm M(\bar \omega \Delta t) =  
\begin{bmatrix}
\cos(\bar \omega \Delta t) & - \sin(\bar \omega \Delta t) \\
\sin(\bar \omega \Delta t) & \cos(\bar \omega \Delta t)
\end{bmatrix}.
\end{equation}
Inserting Eqs.~(\ref{eq:B5})-(\ref{eq:B6}) in Eq.~(\ref{eq:rfullv2_mat}), we get
\begin{align}
    \bm r(t+\Delta t) &= \bm r(t) + \bar{\bm v}_\text{c} \Delta t \notag +\left[\rho_1A_1 \bm {\hat n_1}(t)+\rho_2A_2 \bm {\hat n_2}(t)\right] \nonumber \\ & \ \ \ - \bm M(\bar \omega \Delta t)\left[\rho_1A_1 \bm {\hat n_1}(t)+\rho_2A_2 \bm {\hat n_2}(t)\right].
    \label{eq:rfullv2_mat2}
\end{align}
Replacing the terms in the square brackets with $\bm r_\text{c}(t)-\bm r(t)$ from Eq.~(4) in the main manuscript and simplifying, we obtain
\begin{align}
    \bm r(t+\Delta t) = \bm M(\bar \omega \Delta t)\left[\bm r(t)-\bm r_\text{c}(t)\right]+\bm r_\text{c}(t) + \bar{\bm v}_\text{c} \Delta t.
\end{align}
Writing this equation out for two consecutive time steps $t = k\Delta t$ and $t + \Delta t = (k+1) \Delta t$ yields
\begin{align} 
   \bm r^{(k+1)} = \bm M(\bar \omega^{(k)} \Delta t) (\bm r^{(k)}-\bm r_\text{c}^{(k)}) + \bm r_\text{c}^{(k)} + \bm {\bar v}_\text{c} ^{(k)}\Delta t.
\end{align}
We allow $\rho_2$ to change from one time step to another. This requires $\bm r_\text{c}$ to be recalculated, as detailed in Algorithm~1 in the main manuscript.

\newpage
\section{Comparison of experimental and simulated trajectories for different numbers of Fourier modes included}\label{app:C}

\begin{figure}[!htb]
\includegraphics[width=0.44\textwidth]{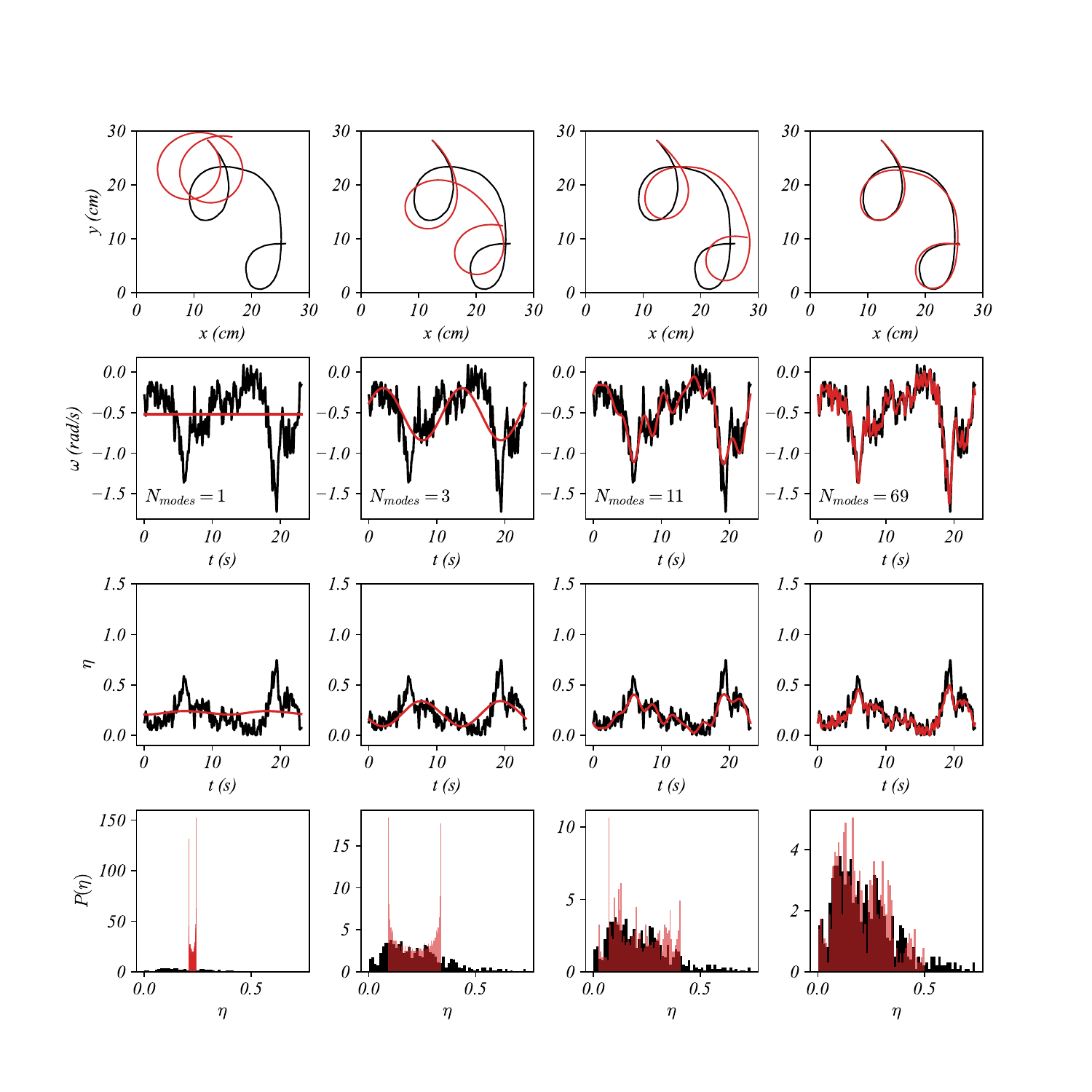}
\caption{The top row shows synthetic trajectories corresponding to the experimental trajectory in the third column of Fig.~3 in the main manuscript, for different values of $\omega$ picked in the simulations. These $\omega$-values are picked by first plotting $\omega$ vs time (black curves in the second row), and then taking Fourier modes to different orders (red curves in the second row). The number of Fourier modes used is, respectively, $1$, $3$, $11$, and $69$ from left to right in the second row. The third row shows the resulting variation of $\eta$ with time, as measured in the experiments (in black) and in the simulations (in red). The bottom row shows the histograms for $\eta$ values, as found for the experiments (in black) and from the simulations (in red).}
\label{fig:fourier} 
\end{figure}

\newpage

\newpage

\section{Cost matrices for TinyMPC} \label{app:D}

Before implementing the TinyMPC pipeline on a physical brainbot, simulations were carried out on a virtual bot to validate the control framework. These simulations focused on solving a similar lemniscate tracking problem, replacing the external positioning setup in Fig. 5 of the main manuscript with a simulation-based state measurement. One of the objectives of this exercise was to fine-tune the weight matrices $Q$ and $R$, associated with the tracking cost and the control effect, respectively. The weights contained in these matrices are mentioned in Eq.~(\ref{eq:cost_Q_R}), and have been implemented in the experimental implementation of the control pipeline,
\begin{equation}
Q = 
\begin{bmatrix}
10 & 0  & 0 & 0 \\
0 & 10  & 0 & 0 \\
0 & 0  & 1 & 0 \\
0 & 0  & 0 & 1
\end{bmatrix},
\qquad
R =
\begin{bmatrix}
0.1 & 0 \\
0 & 0.1
\end{bmatrix}.
\label{eq:cost_Q_R}
\end{equation}
The costs for the deviation of position and velocity are penalized by weights equal to $10$ and $1$, respectively, while the weights associated with the acceleration (control input) are $0.1$.

The matrix $P$, concerning the weights associated with the terminal cost, is computed internally by TinyMPC using the Riccati recursion method [28]. The weights contained in this matrix are delineated in Eq.~(\ref{eq:cost_P}),
\begin{equation}
    P = 
    \begin{bmatrix}
    131.3201 & 0 & 50.2887 & 0 \\
    0 & 131.3201 & 0 & 50.2887 \\
    50.2887 & 0 & 54.0187 & 0 \\
    0 & 50.2887 & 0 & 54.0187
    \end{bmatrix}.
    \label{eq:cost_P}
\end{equation}

\end{document}